\documentclass[conference]{IEEEtran}
\IEEEoverridecommandlockouts
\usepackage{cite}
\usepackage{amsmath,amssymb,amsfonts}
\usepackage{algorithmic}
\usepackage{algorithm}
\usepackage{graphicx}
\usepackage{textcomp}
\usepackage{xcolor}
\usepackage{tikz}
\usepackage{subfigure}

\newcommand{\ddtheta}{\frac{d}{d\theta}}

\newcommand{\thetai}{{(\theta,i)}}

\usepackage{hyperref}

\usetikzlibrary{shapes, arrows, positioning, calc, fit}

\def\BibTeX{{\rm B\kern-.05em{\sc i\kern-.025em b}\kern-.08em
    T\kern-.1667em\lower.7ex\hbox{E}\kern-.125emX}}
    
\newcommand\blfootnote[1]{%
  \begingroup
  \renewcommand\thefootnote{}\footnote{#1}%
  \addtocounter{footnote}{-1}%
  \endgroup
}
    
\makeatletter
\newcommand{\linebreakand}{%
  \end{@IEEEauthorhalign}
  \hfill\mbox{}\par
  \mbox{}\hfill\begin{@IEEEauthorhalign}
}
\makeatother
\begin{document}

\title{Inference of Stochastic Disease Transmission Models Using Particle-MCMC and a Gradient Based Proposal \thanks{CR was supported by a Research Studentship jointly funded by EPSRC and the ESRC Centre for Doctoral Training on Quantification and Management of Risk and Uncertainty in Complex Systems Environments [EP/L015927/1]; SM was supported by the EPSRC through the Big Hypotheses grant [EP/R018537/1] and CR and JPG were supported by funding from the UK Health Security Agency (UKHSA).}}

\author{\IEEEauthorblockN{Conor Rosato}
\IEEEauthorblockA{\textit{Dept. of Electrical Engineering and Electronics} \\
\textit{University of Liverpool, United Kingdom}\\
c.m.rosato@liverpool.ac.uk}
\and
\IEEEauthorblockN{John Harris}
\IEEEauthorblockA{\textit{Public Health England}\\
\textit{United Kingdom}\\
john.p.harris@phe.gov.uk}
\linebreakand 
\IEEEauthorblockN{Jasmina Panovska-Griffiths}
\IEEEauthorblockA{\textit{The Big Data Institute} \\
\textit{The Queens's College, University of Oxford, United Kingdom}\\
jasmina.panovska-griffiths@queens.ox.ac.uk}
\and
\IEEEauthorblockN{Simon Maskell}
\IEEEauthorblockA{\textit{Dept. of Electrical Engineering and Electronics}\\
\textit{University of Liverpool, United Kingdom}\\
s.maskell@liverpool.ac.uk}
}

\maketitle

\begin{abstract}
State-space models have been widely used to model the dynamics of communicable diseases in populations of interest by fitting to time-series data. Particle filters have enabled these models to incorporate stochasticity and so can better reflect the true nature of population behaviours. Relevant parameters such as the spread of the disease, $R_t$, and recovery rates can be inferred using Particle MCMC. The standard method uses a Metropolis-Hastings random-walk proposal which can struggle to reach the stationary distribution in a reasonable time when there are multiple parameters.

In this paper we obtain full Bayesian parameter estimations using gradient information and the No U-Turn Sampler (NUTS) when proposing new parameters of stochastic non-linear Susceptible-Exposed-Infected-Recovered (SEIR) and SIR models. Although NUTS makes more than one target evaluation per iteration, we show that it can provide more accurate estimates in a shorter run time than Metropolis-Hastings.
\end{abstract}


\section{Introduction}
Calibration of epidemiological models using disease specific data is important when striving to understand how a disease evolves with time. Inferring parameters such as the effective contact and recovery rates, especially at the beginning of an outbreak, is vital for public health officials when assessing the deadliness of a disease.\blfootnote{\bf{© 2022 IEEE.  Personal use of this material is permitted.  Permission from IEEE must be obtained for all other uses, in any current or future media, including reprinting/republishing this material for advertising or promotional purposes, creating new collective works, for resale or redistribution to servers or lists, or reuse of any copyrighted component of this work in other works.}}

Two methods for modelling diseases include: Agent-based models (ABMs) \cite{kerr2021covasim} which simulate interactions with individuals within the population based on a set of rules; and the  compartmental approach of splitting the population into unobservable compartments, for example Susceptible, Infected and Recovered (SIR) \cite{kermack1927contribution} and allowing a fraction at every timestep $t$, to progress to the next compartment. State space models (SSMs) are a favoured method for representing the dependence of latent states in non-linear dynamical systems in a wide range of research fields \cite{doucet2001sequential}. A SSM consists of a state equation, 
\begin{equation}\label{xt}
x_{t} \mid x_{t-1} \sim f(x_{t} \mid x_{t-1}, \theta),
\end{equation}
\noindent which is parameterised by $\theta$ and models how the dynamical system moves from the previous state to the current state, $x_t$, at time $t$. They also include an observation equation 
\begin{equation}\label{yt}
y_{t} \mid x_{t} \sim f(y_{t} \mid x_{t}, \theta),
\end{equation}
\noindent which describes how the observation, $y_t$, is linked to the state equation. Particle filters \cite{gordon1993novel, arulampalam_maskell_gordon_clapp_2002} are a popular method for inferring the time-dependent hidden states of SSMs and have been previously applied to compartmental disease transmission models \cite{dukic2012tracking, dawson2015detecting, sheinson2014comparison, shahtori2016sequential}.

The two overarching methods used for estimating parameters in the context of epidemiological models, as outlined in a recent systematic review \cite{Review_calibration}, are optimisation and sampling algorithms. Optimisation algorithms include grid search and iterative, descent-guided optimisation while sampling methods include Markov Chain Monte Carlo (MCMC) techniques, particle-MCMC (p-MCMC), Approximate Bayesian Computation (ABC) and history matching. One difference between these two groups of techniques is that optimisation algorithms find point estimates of the parameters while sampling methods can produce full Bayesian parameter estimates which in-turn will capture parameter uncertainty.

MCMC algorithms work by drawing a sample, which is based on the previous sample, from a target distribution, $\pi(\theta)$. An accept/reject step is included which determines whether to keep or discard the drawn sample. Metropolis-Hastings (MH) randomly samples from $\pi(\theta)$ without reference to its structure. It can therefore take the Markov Chain a long time to reach its stationary distribution if $\theta$ has multiple dimensions. Hamiltonian Monte Carlo (HMC) \cite{neal2012mcmc} and the No-U-Turn Sampler (NUTS) \cite{hoffman2014no} use gradient information about the log-posterior of $\theta$ so can efficiently explore $\pi(\theta)$. Probabilistic programming languages (ppls), such as Stan \cite{carpenter2017stan}, have made MCMC an accessible tool for calibrating epidemiological models \cite{chatzilena2019contemporary, moore2021refining} because these ppls use NUTS. 

Particle-MCMC (p-MCMC) \cite{andrieu_doucet_holenstein_2010} combines the particle filter and MCMC to make state and parameter estimates by performing Bayesian inference on non-linear non-Gaussian scenarios where standard MCMC methods can fail. A detailed explanation is given in section \ref{sec:particle_mcmc}. Particle-MCMC has been used to infer parameters of simulated genealogies and time series data \cite{rasmussen2011inference}, a dengue outbreak \cite{wigren2019parameter}, non-communicable diseases in Egypt \cite{NonCommunicablePMCMC}, cholera in Bangladesh \cite{koepke2016predictive}, COVID-19 in the United Kingdom (UK) \cite{knock2021key}, the 2009 A/H1N1 pandemic in England \cite{dureau2013capturing} and an ABM in \cite{lux2021bayesian}. It has been suggested in \cite{baguelin2020tooling, endo2019introduction} that a potential reason why p-MCMC has yet to become a mainstream method for modelling epidemics is due to the complexity of the mathematics behind the algorithm:  \cite{endo2019introduction} make the methodology more accessible by providing  a step-by-step tutorial. The standard proposal used in calibration of epidemiological models when using p-MCMC is the MH random walk \cite{rasmussen2011inference, NonCommunicablePMCMC, koepke2016predictive, knock2021key, dureau2013capturing, lux2021bayesian, endo2019introduction} which inherits the same drawbacks as explained previously for MCMC. 

In this piece of work we focus on estimating the parameters of stochastic disease compartmental models using the recent development of a variant of p-MCMC that uses NUTS as the proposal\cite{rosato2021efficient}. 

The paper is organised as follows: section~\ref{sec:particle_mcmc} explains how to combine p-MCMC and NUTS in more detail, before section~\ref{sec:model} describes the models we will consider and the results obtained. Section~\ref{sec:conclusions} then concludes the paper.

\section{Particle-MCMC}\label{sec:particle_mcmc}

In this section we outline how to construct the p-MCMC algorithm and show how an unbiased estimate of the log-likelihood, given by the particle filter, can be used in the MH proposal within the MCMC algorithm. We also show how to estimate the gradient of the log-likelihood which is used when proposing a new vector of $\theta'$ using NUTS.


Bayesian calibration of SSMs seen in (\ref{xt}) and (\ref{yt}) is undertaken with the goal of estimating the parameter posterior distribution $p(\theta|y)$, which involves finding a set of $\theta$ that best represents the data, $y$. This can be done by using Bayes theorem:

\begin{equation}
p(\theta|y)=\frac{p(y|\theta) p(\theta)}{p(y)},
\end{equation}

\noindent where $p(\theta)$ is the prior.

\subsection{Metropolis Hastings}\label{sec:MH}

In the well understood MH proposal a set of parameters $\theta'$ is drawn from a proposal distribution $q(\theta'|\theta)$, which is commonly chosen to be a Gaussian centered around the current position, $\theta$, and a variance parameter, also known as step-size, which is chosen by the user. An accept or reject step is introduced that determines whether to accept $\theta'$ as part of the Markov chain or reject and revert back to $\theta$. If $\theta'$ is accepted the inputs for the next iteration are $\theta$ = $\theta'$ and $p(y_{1:T}|\theta)$ = $p(y_{1:T}|\theta')$. However if $\theta'$ is rejected the inputs are $\theta$ = $\theta$ and $p(y_{1:T}|\theta)$ = $p(y_{1:T}|\theta)$. This process occurs for, $M$ MCMC iterations and at each iteration two values are stored: a current value of $\theta$ and a particle filter estimate of $p(y_{1:T}|\theta)$. The acceptance probability is
\begin{equation}
\alpha(\theta,\theta') = \min \left\{1, \frac{\pi(\theta')}{\pi(\theta)} \frac{q(\theta |  \theta')}{q(\theta' |  \theta)}\right\}.
\label{eq:MHalgorithm}
\end{equation}

\noindent Note $\pi(\theta)$ = $p(\theta|y)$. When estimating $\pi(\theta)$ a tractable likelihood is needed which in many cases is not available. The particle filter (explained in section \ref{sec:particleFilter}) provides a method to approximate the marginal likelihood, $p(y_{1:T}|\theta)$ which allows us to formulate the acceptance step as

\begin{equation}
\alpha(\theta,\theta') = \min \left\{1, \frac{p\left(\theta^{\prime}\right)}{p(\theta)} \frac{p(y_{1:T}|\theta')}{p(y_{1:T}|\theta)} \frac{q(\theta |  \theta')}{q(\theta' | \theta)}\right\}.
\label{eq:MHalgorithm_}
\end{equation}

\noindent This is the particle marginal Metropolis-Hastings (PMMH) algorithm described in \cite{andrieu_doucet_holenstein_2010} and is described in Algorithm~\ref{p-mcmcMH}.

\subsection{No-U-Turn Sampler}\label{sec:NUTS}

NUTS \cite{hoffman2014no} is a gradient based method and an extension of HMC \cite{duane1987hybrid, betancourt2017conceptual}. Both methods use the leapfrog numerical integrator to generate new samples by discretising  Hamilton’s equations which are defined to be

\begin{align}
\frac{d \theta}{d t}&=\frac{\partial H}{\partial m}\nonumber \\\\
\frac{d m}{d t}&=-\frac{\partial H}{\partial \theta},
\nonumber
\end{align}

\noindent where $m$ is the momentum. The Hamiltonian $H$ is defined to be \begin{align}
H(\theta,m) = U(\theta) + K(m),
\end{align}

\noindent where $K(m)$ is the kinetic energy and $U(\theta)$ is the potential energy. Hamilton's equations are a pair of differential equations that describe a system in terms of its position and momentum where the potential of the system is defined by $U=-\log(\pi(\theta))$. Sampling a random initial $m$, from a Gaussian, a new state $\theta'$ is proposed using the gradient of $\pi(\theta)$. The leapfrog method is symplectic, meaning it preserves the geometry of the system which leads to high acceptance rates, and reversible, meaning detailed balance is maintained. The generated samples from HMC are governed by two user defined parameters: the number of steps $L$ and the stepsize used by leapfrog $\Delta$. NUTS adaptively calibrates $L$ by building trajectories that take steps both forward and backwards, doubling in number at each iteration. This is done until the trajectory starts to double back on itself. It is at this point that a state is sampled from a complete balanced tree. The result is a reversible sampling process that maintains detailed balance and is described in Algorithm 3.

\subsection{Particle Filter}\label{sec:particleFilter}

The benefit of using a particle filter is its ability to model SSMs that are nonlinear and non-Gaussian. It does this by representing the probability density function over the hidden states, $x_t$, with a set of $N$ random samples (particles), each being an independent hypothesis of $x_t$. 

The particle filter propagates particles forward in time using the model function (often referred to as the dynamics) in (\ref{xt}) which encodes some prior information on how the state evolves with time. Due to $\theta$ being static, some noise needs to be introduced which will improve the diversity among the drawn samples. An observation model (often referred to as the likelihood), seen in (\ref{yt}), is used to associate a measurement with the state, $x_t$, at every timestep. The dynamics and likelihood can be parameterised by $\theta$ such that

\begin{align}\label{dynamicsLikelihood}
p\left(y_{1:t},x_{1:t}|\theta\right) = p(y_1&|x_1,\theta)p\left(x_1|\theta\right) \\ \nonumber
&\times \prod_{\tau=2}^{t}p\left(y_\tau|x_\tau,\theta\right)p\left(x_\tau|x_{\tau-1},\theta\right).
\end{align}

\noindent The $i$th particle has a state sequence, $x_{1:t}^{(\theta, i)}$, which is modelled using a Markov chain to describe the current state at time, $t$, only being dependent on the previous state at time, $t-1$.
The $i${th} sample has an associated importance weight, $w_t^{(\theta, i)}$, which is determined by how probable it is given the dynamics and new measurement. Broadly speaking, large weights correspond to more accurate representations of the dynamical system and lower weights the opposite. A set of $N$ particles can then be represented as $\left\{x_{1:t}^{(\theta, i)}, w_t^{(\theta, i)}\right\}_{i=1}^{N}$.

At $t=0$, weights are set to be $1/N$ but are recursively calculated at every timestep when $t>0$ using 

\begin{align}\label{weightupdate}
w_{1:t}^{(\theta,i)}=w_{1:t-1}^{(\theta,i)}\frac{p\left(y_t|x_t^{(\theta,i)}\right)p\left(x_t^{(\theta,i)}|x^{(\theta,i)}_{t-1},\theta\right)}{q\left(x_t^{(\theta,i)}|x_{t-1}^{(\theta,i)},y_t\right)},
\end{align}

\noindent where ${q\left(x_t^{(\theta,i)}|x_{t-1}^{(\theta,i)},y_t\right)}$ is the proposal. Different options exist for formulating a proposal which are described in \cite{rosato2021efficient}. However, in this paper we choose the proposal to be equal the dynamic model, $q\left(x_t^{(\theta,i)}|x_{t-1}^{(\theta,i)},y_t\right)=p\left(x_t^{(i)}|x^{(\theta,i)}_{t-1},\theta\right)$, which simplifies (\ref{weightupdate}) to 

\begin{align}
	w_{1:t}^{(\theta,i)}=p\left(y_t|x_t^{(\theta,i)}\right)w_{1:t-1}^{(\theta,i)}.
	\label{eq:priorproposalweight}
\end{align}



\noindent The joint distribution, $p\left(y_{1:t},x_{1:t}|\theta\right)$, can be estimated using the unnormalised weights from (\ref{eq:priorproposalweight}) such that

\begin{align}
\int p\left(y_{1:t},x_{1:t}|\theta\right)f\left(x_{1:t}\right) dx_{1:t} \approx \frac{1}{N}\sum_{i=1}^N w_{1:t}^{(\theta,i)}f\left(x_{1:t}^{(i)}\right),\label{eq:jointexp}
\end{align}

\noindent which is unbiased. The posterior distribution, $p\left(x_{1:t}|y_{1:t},\theta\right)$, can then be estimated by 

\begin{equation}
\int p\left(x_{1:t}|y_{1:t},\theta\right)f\left(x_{1:t}\right) dx_{1:t}= \int \frac{p\left(y_{1:t},x_{1:t}|\theta\right)}{p\left(y_{1:t}|\theta\right)}f\left(x_{1:t}\right) dx_{1:t}.
\end{equation}

\noindent If $f\left(x_{1:t}\right)=1$ then 

\begin{equation}
p\left(y_{1:t}|\theta\right) = \int p\left(y_{1:t},x_{1:t}|\theta\right)dx_{1:t}\approx  \frac{1}{N}\sum_{i=1}^N w_{1:t}^{(\theta,i)},\label{eq:likelihood}
\end{equation}

\noindent which is similar to (\ref{eq:jointexp}). The normalised weights (which differ from those in (\ref{eq:priorproposalweight})) can be calculated using 

\begin{equation}
\tilde{w}_{1:t}^{(\theta,i)} = \frac{w_{1:t}^{(\theta,i)}}{\sum_{j=1}^N w_{1:t}^{(\theta,j)}},
\label{eq:normalisedweights}
\end{equation}

\noindent which are used when estimating the integral 

\begin{align}
\int p\left(x_{1:t}|y_{1:t},\theta\right)f\left(x_{1:t}\right)dx_{1:t} = \sum_{i=1}^N \tilde{w}_{1:t}^{(\theta,i)}f\left(x_{1:t}^{(\theta,i)}\right).
\label{eq:a}
\end{align}

\noindent We note that while (\ref{eq:jointexp}) is an unbiased estimate, \ref{eq:a}) is biased (due to it being a ratio of estimates).


\subsubsection{Resampling}\label{sec:resampling}

The methods described in section (\ref{sec:particleFilter}) until now results in the sequential importance sampling (SIS) algorithm. One limitation of SIS is that the normalised weights in (\ref{eq:normalisedweights}) becoming increasingly skewed at each increment of $t$ with one becoming close to 1 and the others 0. 

A widely used method for overcoming this is to calculate the number of effective samples at each iteration of $t$ as 

\begin{equation}
N_{\text{eff}} = \frac{1}{\sum_{i=1}^N \left(\tilde{w}_{1:t}^{(\theta,i)}\right)^2},
\label{eq:neff}
\end{equation}

\noindent and comparing $N_{eff}$ with a user-defined threshold. If $N_{eff}$ is less than the threshold, resampling is implemented. Multinomial resampling is a popular method which involves drawing from the current particle set $N$ times, with probabilities proportional to the corresponding normalised weights. This allows for particles that have a better guess of $x_t$ and higher weights to get replicated while particles with lower weights die off. After resampling the normalised and unnormalised weights are set to $\frac{1}{N}\sum_{i=1}^N w_{1:t}^{(\theta,i)}$ and $1/N$, respectively. 

\subsection{Calculating Likelihood and Gradients}

In this section we outline how to calculate the likelihood and the gradient of the likelihood w.r.t $\theta$. 

\subsubsection{Likelihood}\label{sec:likelihood}

The likelihood can be calculated from (\ref{eq:likelihood}) which is the sum of the unnormalised weights when $t=T$. It is a byproduct of running the particle filter so no other calculations need to be undertaken.

\subsubsection{Gradients}\label{sec:gradients}

As previously described, MH randomly samples $\theta'$ from a proposal distribution $q(\theta'|\theta)$ which is determined by a user-defined step-size. If $\theta$ has multiple dimensions or the correlation between $\theta$ is high, the sampler can take a long time to search $\pi(\theta)$ due to being stuck in local maximum peaks. It can also be time consuming to select a suitable step-size for $\theta$ as choosing one that results in large jumps can mean $\theta'$ rarely gets accepted while having too small a step-size may result in $\pi(\theta)$ being poorly explored. This problem is accentuated if $\theta$ is high dimensional. The work done in \cite{rosato2021efficient} shows how to differentiate the particle filter so gradients of the log-posterior of $\theta$ can be used to efficiently propose $\theta'$. 

We present the gradient of the log-posterior of $\theta$ as
\begin{align}
\nabla \log p(\theta|y_{1:t})=\nabla \log p(\theta)+\nabla \log p(y_{1:t}|\theta),
\label{eq:gradientlogposterior}
\end{align}
where $\nabla \log p(\theta)$ is the gradient of the log-prior and $\nabla \log p(y_{1:t}|\theta)$ is the gradient of the log-likelihood\footnote{Working with logs is often found to be more numerically stable.}. Note that we assume $\nabla \log p(\theta)$ can be calculated explicitly.

Differentiating the weights gives an approximation to the gradient of the likelihood:
\begin{eqnarray}
	\ddtheta p(y_{1:t} | \theta) & = & \frac{1}{N} \sum_{i=1}^N \ddtheta w_{1:t}^{(\theta,i)}. \label{eqn:dlikweightsum}
\end{eqnarray}

Applying the Chain Rule to (\ref{eq:likelihood}) and (\ref{eqn:dlikweightsum}) gives
\begin{eqnarray}
	\ddtheta \log p(y_{1:t} | \theta) & = & \frac{1}{p(y_{1:t} | \theta)}\sum_{i=1}^N w_{1:t}^{(\theta,i)}\ddtheta \log w_{1:t}^{(\theta,i)}
		\\ & \approx &
	\sum_{i=1}^N \tilde{w}_{1:t}^{(\theta,i)}\ddtheta \log w_{1:t}^{(\theta,i)}	\label{eqn:dloglik}
\end{eqnarray}

\noindent In the general case we need to calculate the derivatives of the dynamics, ${\frac{d}{d\theta}p\left(x_t^{(\theta,i)}|x^{(\theta,i)}_{t-1},\theta\right)}$, likelihood, $\frac{d}{d\theta}p\left(y_t|x_t^{(\theta,i)}\right)$, and proposal, ${\frac{d}{d\theta}q\left(x_t^{(\theta,i)}|x_{t-1}^{(\theta,i)},y_t\right)}$ which are described in detail in \cite{rosato2021efficient}. However, similarly to (\ref{eq:priorproposalweight}), because we set the proposal to equal the dynamic model, the derivatives of the weights simplifies to 

\begin{align}\label{simplifieddweightupdatedtheta}
\frac{d}{d\theta}w_{1:t}^{(\theta,i)}=\frac{d}{d\theta}p\left(y_t|x_t^{(\theta,i)}\right)w_{1:t-1}^{(\theta,i)}+p\left(y_t|x_t^{(\theta,i)}\right)\frac{d}{d\theta}w_{1:t-1}^{(\theta,i)}.
\end{align}

\noindent Therefore in this paper we will only describe how to calculate (\ref{simplifieddweightupdatedtheta}). 

\paragraph{Derivatives of Particle States}\label{sec:particlestates}

It is evident that $\frac{d}{d\theta}p\left(y_t|x_t^{(\theta,i)}\right)$ does not explicitly depend on $\theta$ but $x_t^{(\theta,i)}$ does. Therefore in order for the derivative of the likelihood w.r.t $\theta$ to not equal zero we firstly need to calculate $dx_{t}^{(\theta, i)}/d\theta$. If we define 

\begin{align}
x_k^{(\theta, i)} = f(x_{t-1}^{(\theta, i)}, \theta, y_t, \epsilon^i_t), 
\end{align}

\noindent then the derivative can be calculated by

\begin{eqnarray}
	\frac{dx_t^{(\theta, i)}}{d\theta} & = & \frac{d}{d\theta}f(x_{t-1}^{(\theta, i)}, \theta, y_t, \epsilon^i_t) \label{eqn:df} \\ & = &
		\frac{\partial f}{\partial x_{t-1}^{(\theta, i)}}\frac{d x_{t-1}^{(\theta, i)}}{d \theta} + \frac{\partial f}{\partial\theta} \label{eqn:partialf}
		\frac{d \theta}{d \theta} \\ & = & 
		\frac{\partial f}{\partial x_{t-1}^{(\theta, i)}}\frac{d x_{t-1}^{(\theta, i)}}{d \theta} + \frac{\partial f}{\partial\theta}.	\label{eqn:dxk}
\end{eqnarray}

\paragraph{Derivatives of Gaussian Likelihood}\label{sec:gaussianlike}

We now describe how to calculate the derivative of a Gaussian likelihood where
\begin{align}
    	L(x^{(\theta, i)}_t, \theta, y_t) &\triangleq \log p\left(y_t | x^{(\theta, i)}_t \right) \\ &= \log\mathcal{N}\left(y_t; h(x^{(\theta, i)}_t, \theta), R(\theta)\right),
\end{align}

and 

\small
\begin{align}
	\frac{d}{d\theta} L\left(x_t^{(\theta, i)}, \theta, y_t\right) =&
		\frac{\partial}{\partial h}\log\mathcal{N}(y_t; h, R)\left(\frac{\partial h}{\partial x^\thetai_t}\frac{d x^\thetai_t}{d \theta} +
			\frac{\partial h}{\partial \theta}\right) \nonumber \\ &+ \frac{\partial}{\partial R}\log\mathcal{N}(y_t; h, R)\frac{d R}{d \theta}
\end{align}
\normalsize

\noindent The derivatives of $\log\mathcal{N}(f; n, R)$ are given in Appendix \ref{app:normalderiv}.

\subsubsection{Common Random Number Resampling}\label{sec:CRNresampling}

It is stated in the literature that the resampling step, described in section (\ref{sec:resampling}), cannot be differentiated. We will not outline here why this is the case but direct the reader to \cite{rosato2021efficient} for a comprehensive explanation. \cite{rosato2021efficient} also outlines how the resampling step can be differentiated when fixing the random number seed and employing multinomial resampling. The approach adopted here does result in discontinuities in an unbiased estimate of the log likelihood but also runs in $O(N)$ time.  

Unlike the standard particle filter described in section (\ref{sec:particleFilter}) we also need to resample:

\begin{itemize}

\item The derivatives of the particle states (see (\ref{eqn:dxk})): $\frac{dx_t^{(\theta, i)}}{d\theta}$.

\item The weight derivatives (see (\ref{simplifieddweightupdatedtheta})) $\frac{d}{d\theta} w^{{(\theta,i)}}_{1:t}$. 
\end{itemize}

\noindent This will ensure that any derivative calculations made after resampling are a function of the parent particle.

\section{Models and Results}\label{sec:model}

In this piece of work we extend the SIR model in \cite{sheinson2014comparison} to include an exposed compartment as shown in section \ref{model:SEIR}. The SEIR model follows a similar discrete time approximation to the ones outlined in \cite{sheinson2014comparison, shahtori2016sequential}. Stochasticity is introduced when modelling the dynamics by including a noise term, $\epsilon_x$, for each time-varying parameter, $x$, which mimics the interactions between people in the population, $P$. Note $\epsilon_x$ will be independent for different values of $x$ and are drawn from $\epsilon_{x} \sim \mathcal{N}(0, \sqrt{\theta} / P)$. An example for $\beta$ would be $\epsilon_{\beta} \sim  \mathcal{N}(0, \sqrt{\beta} / P)$. Table \ref{table:params_and_priors} gives a description of the parameters used in the statistical models below. Prior information for these parameters is also included and taken from \cite{moore2021refining}.

\begin{table*}[]
\centering
\caption{Table of the parameters and a description used in the statistical SIR and SEIR models. Prior information is also included.}
\begin{tabular}{c c c}
\hline 
\bf{Parameter} & \bf{Description} & \bf{Prior Information}\\
\hline 
$P$ & The total population & -\\
$S(t)$ & The proportion of people in the susceptible compartment & $1-(E_0 + I_0)$\\
$E(t)$ & The proportion of people in the exposed compartment & $Unif(0.00016, 0.00024)$\\
$I(t)$ & The proportion of people in the infectious compartment & $Unif(0.00016, 0.00024)$\\
$R(t)$ & The proportion of people in the recovered compartment & -\\
$\beta$ & Mean rate of people an infected person infects per day & $\text{HalfNormal}\left(0.0, 0.5\right)$  \\
$\gamma$ & The proportion of infected recovering per day & $\text{Normal}(4.0, 5.0)$   \\ 
$\delta$ & Length of incubation period & $\text{Normal}(4.0, 5.0)$ \\ 
$R_0$ & The total number of people an infected person infects & - \\
\hline 

\end{tabular}
\label{table:params_and_priors}
\end{table*}

\subsection{SIR Model}\label{model:SIR}

\begin{align}
&s_{t+1}=s_{t}-\beta i_{t} s_{t}^{v}+\epsilon_{\beta}, \label{SIRR:s}\\
&i_{t+1}=i_{t}+\beta i_{t} s_{t}-\gamma \label{SIRR:i} i_{t}-\epsilon_{\beta}+\epsilon_{\gamma}\\
&r_{t+1}=1 - s_{t+1} - i_{t+1},
\end{align}

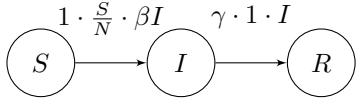
\begin{figure}[!h]
\centering
\begin{tikzpicture}[node distance=0.95cm,auto,>=latex',every node/.append style={align=center},int/.style={draw, circle, minimum size=0.9cm}]
    \node [int] (S) {$S$};
    \node [int, right=of S] (I) {$I$};
    \node [int, right=of I] (R) {$R$};
    
    \path[->, auto=false] (S) edge node {$1 \cdot \frac{S}{N} \cdot \beta I$ \\[2.5em]} (I);
    \path[->, auto=false] (I) edge node {$\gamma \cdot 1 \cdot I$ \\[2.5em]} (R);
\end{tikzpicture}
\caption{A graphical representation of the SIR transmission model.}
\label{fig:SIR:transmission_model}
\end{figure}

\begin{figure}[]
\centering
\subfigure[]{\includegraphics[width=0.48\linewidth]{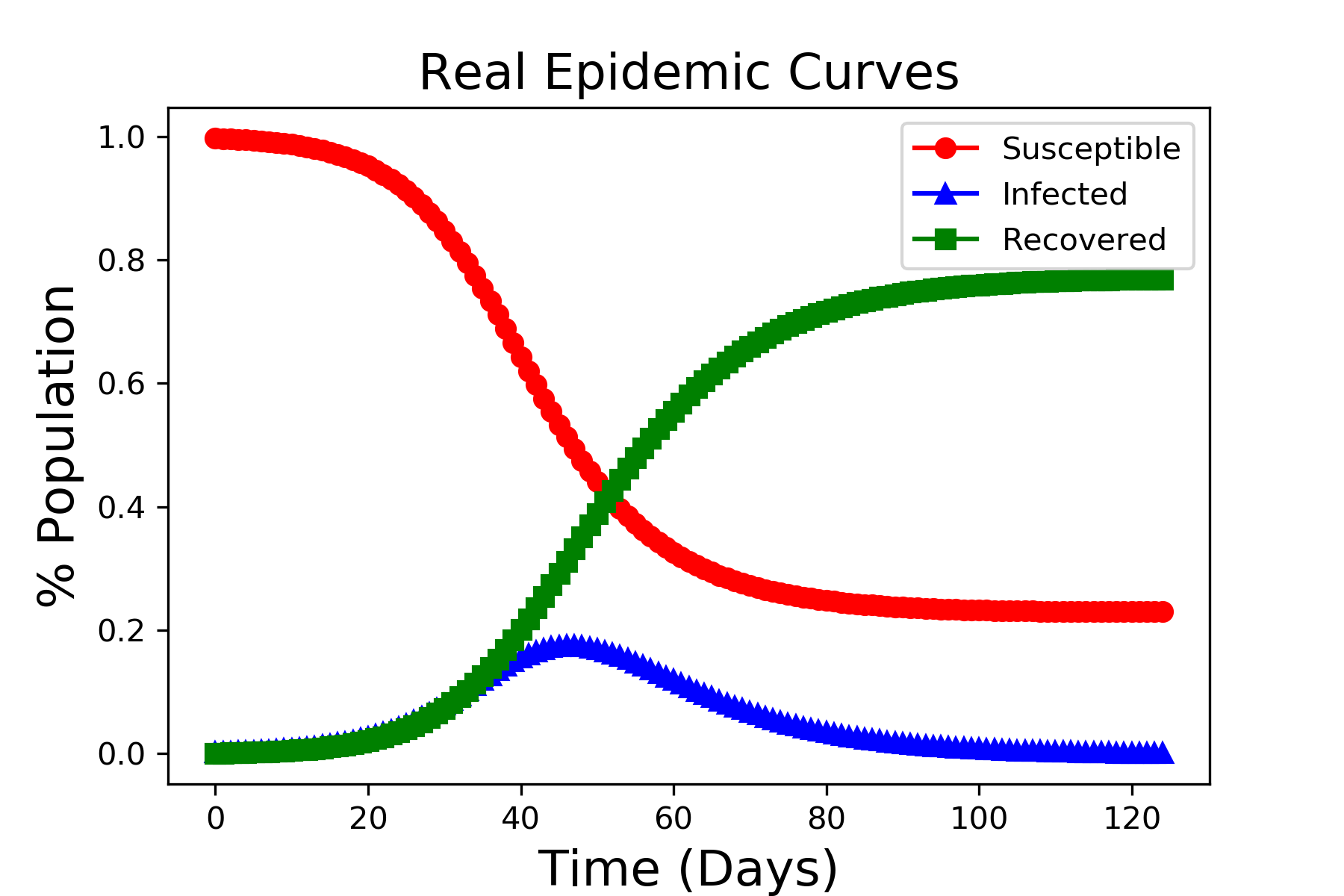}}
\hfill
\subfigure[]{\includegraphics[width=0.48\linewidth]{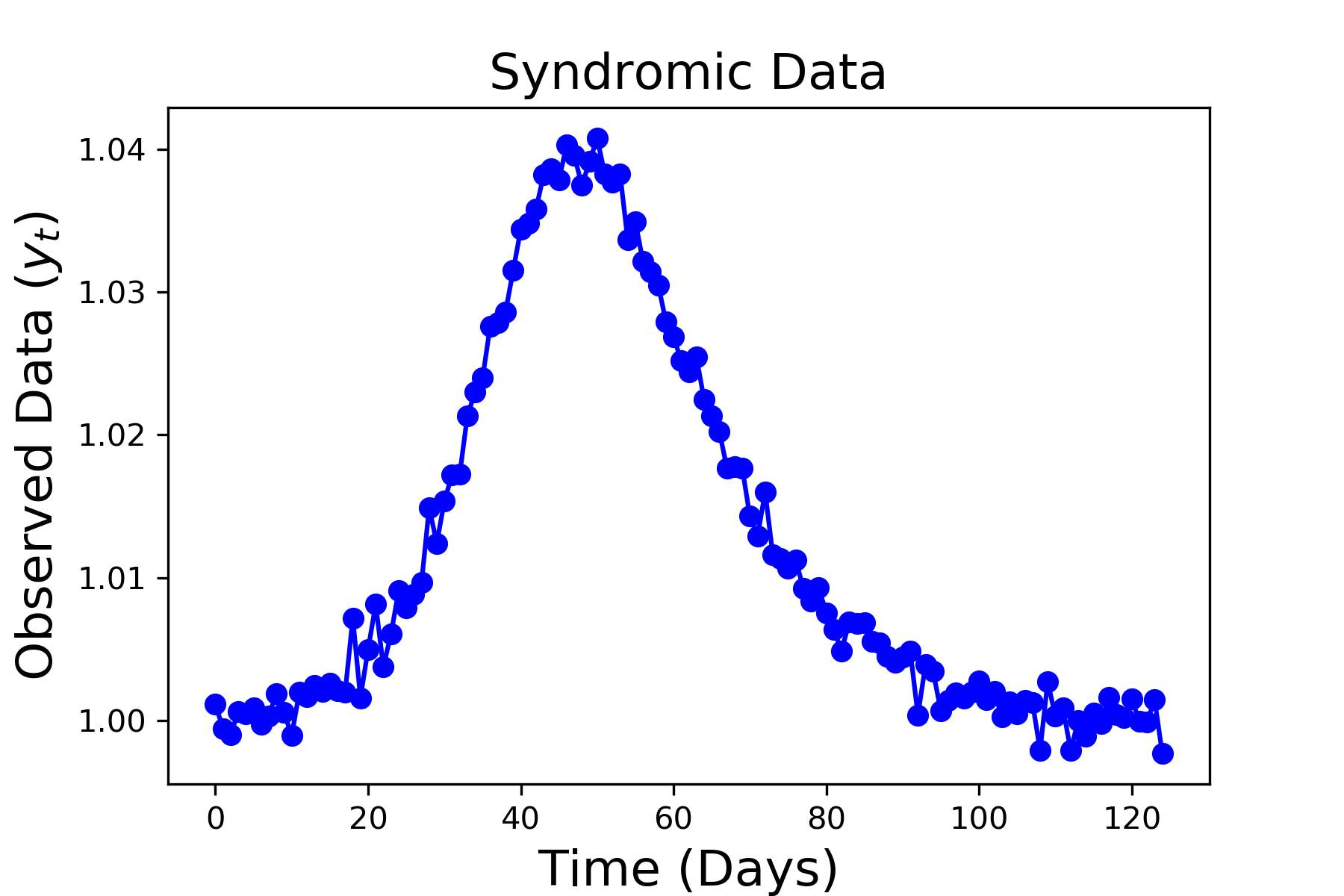}}
\hfill
\subfigure[]{\includegraphics[width=0.48\linewidth]{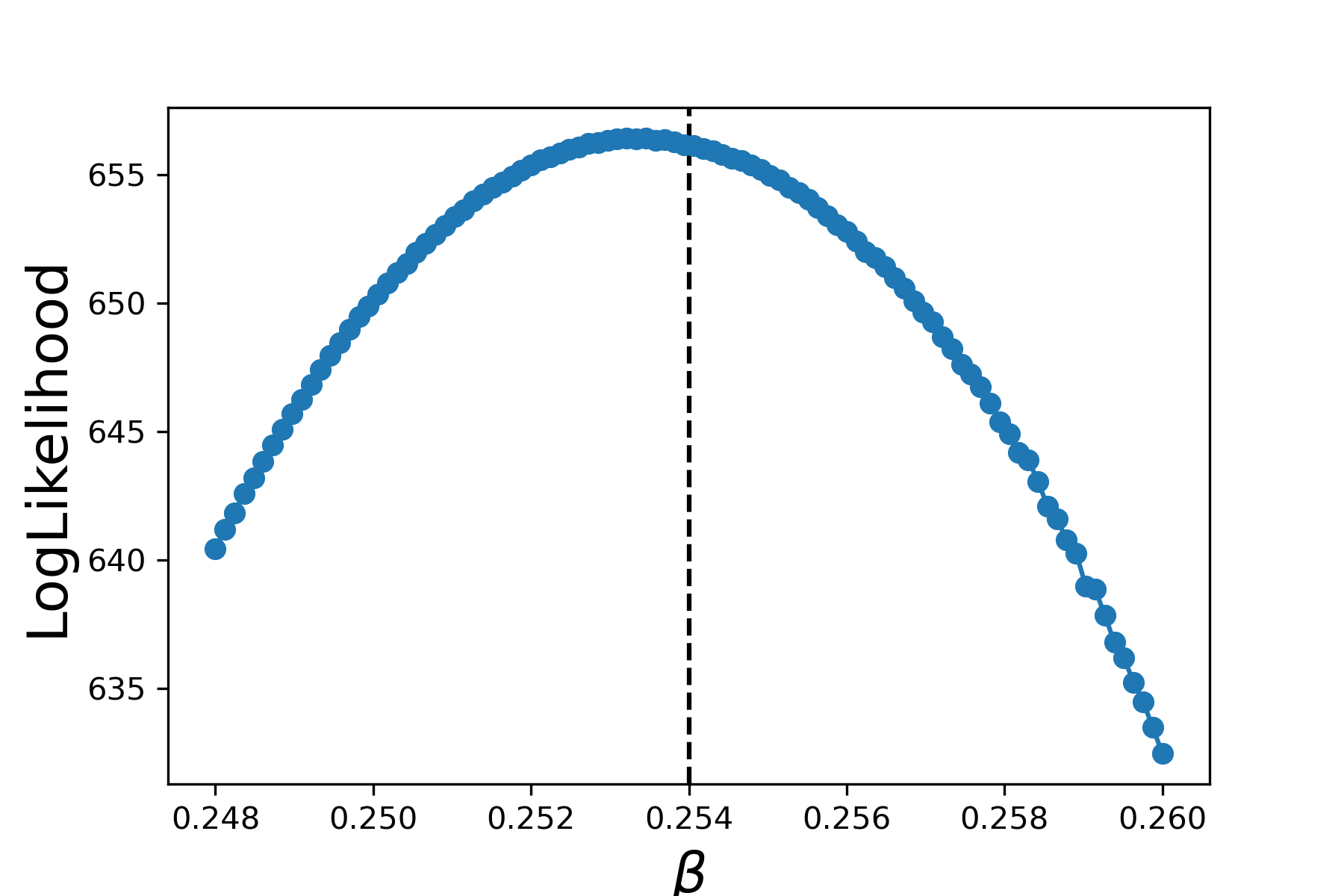}}
\hfill
\subfigure[]{\includegraphics[width=0.48\linewidth]{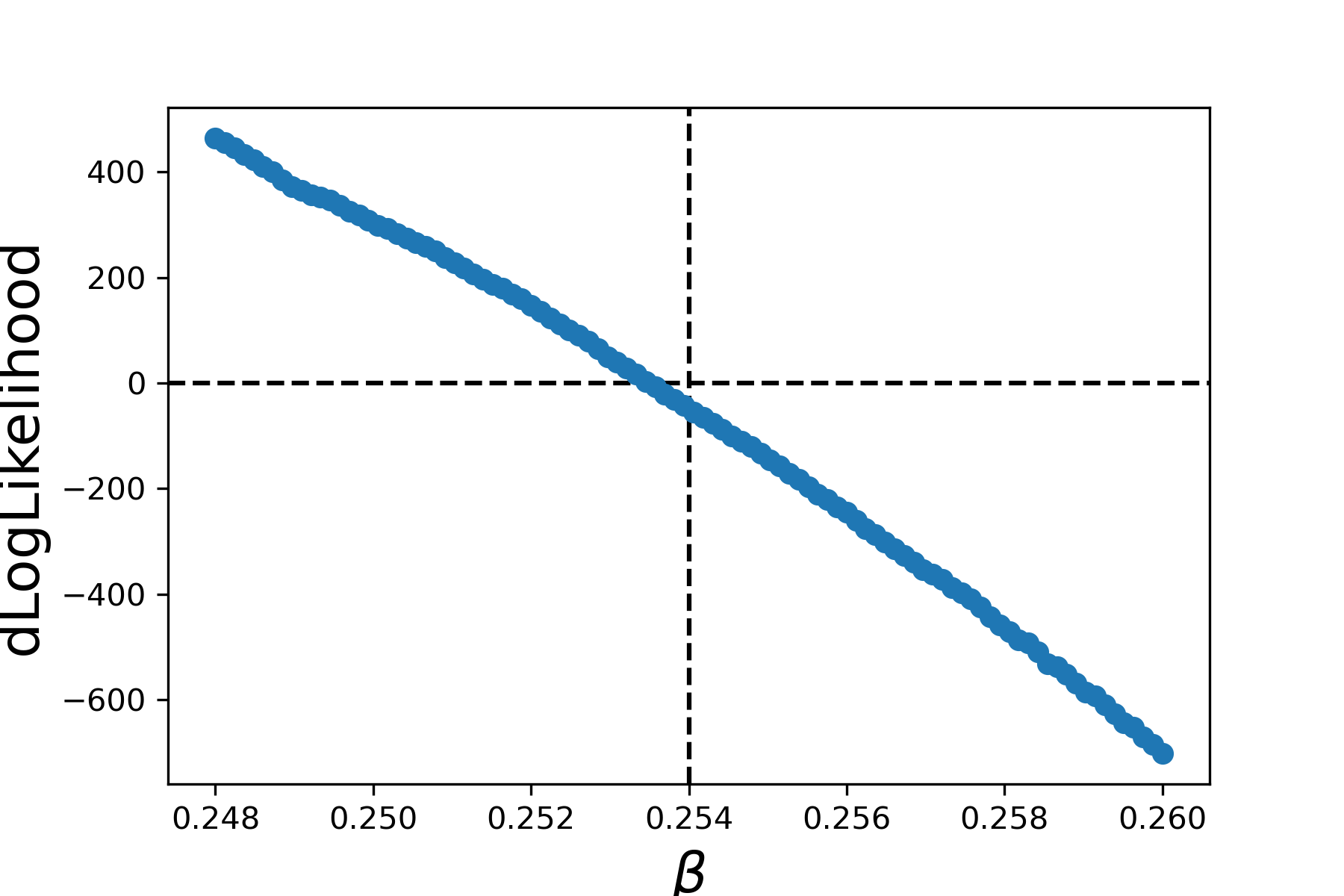}}
\caption{(a) Simulated epidemic curves for the SIR model in section \ref{model:SIR} with $\theta=\left[\beta, \gamma, v\right]$ are set to 0.254, 0.111 and 1.246, respectively, (b) Syndromic data simulated from the Infected curve. The Log-likelihood and the gradient of the log-likelihood across a range of 100 values of $\beta$ equally spaced from 0.248 and 0.260 can be seen in (c) and (d), respectively. The black dashed vertical lines indicate the true value of beta and the horizontal line indicates a gradient of 0.}
\label{EpidemicCurves}
\end{figure}

\subsection{SEIR Model}\label{model:SEIR}

\begin{align}
&s_{t+1}=s_{t}-\beta i_{t} s_{t}+\epsilon_{\beta},\label{SEIR:s} \\
&e_{t+1}=e_{t}+\beta i_{t} s_{t} -\delta e_{t}-\epsilon_{\beta}+\epsilon_{\delta},\label{SEIR:e} \\
&i_{t+1}=i_{t}+\delta e_{t}-\gamma i_{t} +\epsilon_{\gamma}-\epsilon_{\delta},\label{SEIR:i}\\
&r_{t+1}=r_{t}+-\gamma i_{t}+\epsilon_{\gamma}\label{SEIR:r}.
\end{align}

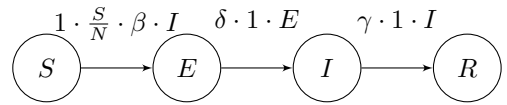
\begin{figure}[!h]
\centering
\begin{tikzpicture}[node distance=0.95cm,auto,>=latex',every node/.append style={align=center},int/.style={draw, circle, minimum size=0.9cm}]
    \node [int] (S) {$S$};
    \node [int, right=of S] (E) {$E$};
    \node [int, right=of E] (I) {$I$};
    \node [int, right=of I] (R) {$R$};
    
    \path[->, auto=false] (S) edge node {$1 \cdot \frac{S}{N} \cdot \beta \cdot I$ \\[2.5em]} (E);
    \path[->, auto=false] (E) edge node {$\delta \cdot 1 \cdot E$ \\[2.5em]} (I);
    \path[->, auto=false] (I) edge node {$\gamma \cdot 1 \cdot I$ \\[2.5em]} (R);
\end{tikzpicture}
\caption{A graphical representation of the SEIR transmission model.}
\label{fig:SEIR:transmission_model}
\end{figure}

\subsection{Observation Equation}\label{Simulated_data}

We model the log of the observations by
\begin{align}\label{obslikelihood}
\log y_{l, t} \sim \mathcal{N}\left(b_{l} i_{t}^{\phi_{t}}, \sigma_{l}^{2}\right),
\end{align}
where $b_{l}$, $\phi_{l}$ and $\sigma_{l}$ are all non-negative constants. For the purpose of this analysis $b_{l}$, $\phi_{l}$ and $\sigma_{l}$ are known which allows us to formulate the likelihood as

\begin{align}\label{LNlikelihood}
p\left(y_t | x^{(\theta, i)}_t \right)=\log\mathcal{N}\left(y_t; b_{l} i_{t}^{\phi_{t}}, \sigma_{l}^{2}\right).
\end{align}

\subsection{Results}

\subsubsection{SIR}\label{results_SIR}

Firstly we compare a NUTS and MH proposal in p-MCMC by running the same simulation with parameter and prior choices as described in \cite{sheinson2014comparison} for the SIR model \ref{model:SIR}. We simulate (\ref{SIRR:s}) and (\ref{SIRR:i}) for $T=125$ days with a population of $P=5000$. The true values of $\theta=\left[\beta, \gamma, v\right]$ are set to 0.254, 0.111 and 1.246, respectively. At $t_0$ we set the number of people that are in the susceptible compartment to 4990 and the infected to 10 which results in the fractions of 4990/5000 and 10/5000, respectively. Figure \ref{EpidemicCurves}(a) shows the percentage of the population in each compartment throughout the duration of the epidemic. Figure \ref{EpidemicCurves}(b) shows the simulated syndromic observations, $y_{1:T}$, which are a fraction of the infected compartment. They are drawn from (\ref{obslikelihood}) with parameters $b_{l}$, $\phi_{l}$ and $\sigma_{l}$ set to 0.25, 1.07 and 0.0012, respectively.

\begin{figure}[htp]
\centering
\subfigure[]{\includegraphics[width=0.48\linewidth]{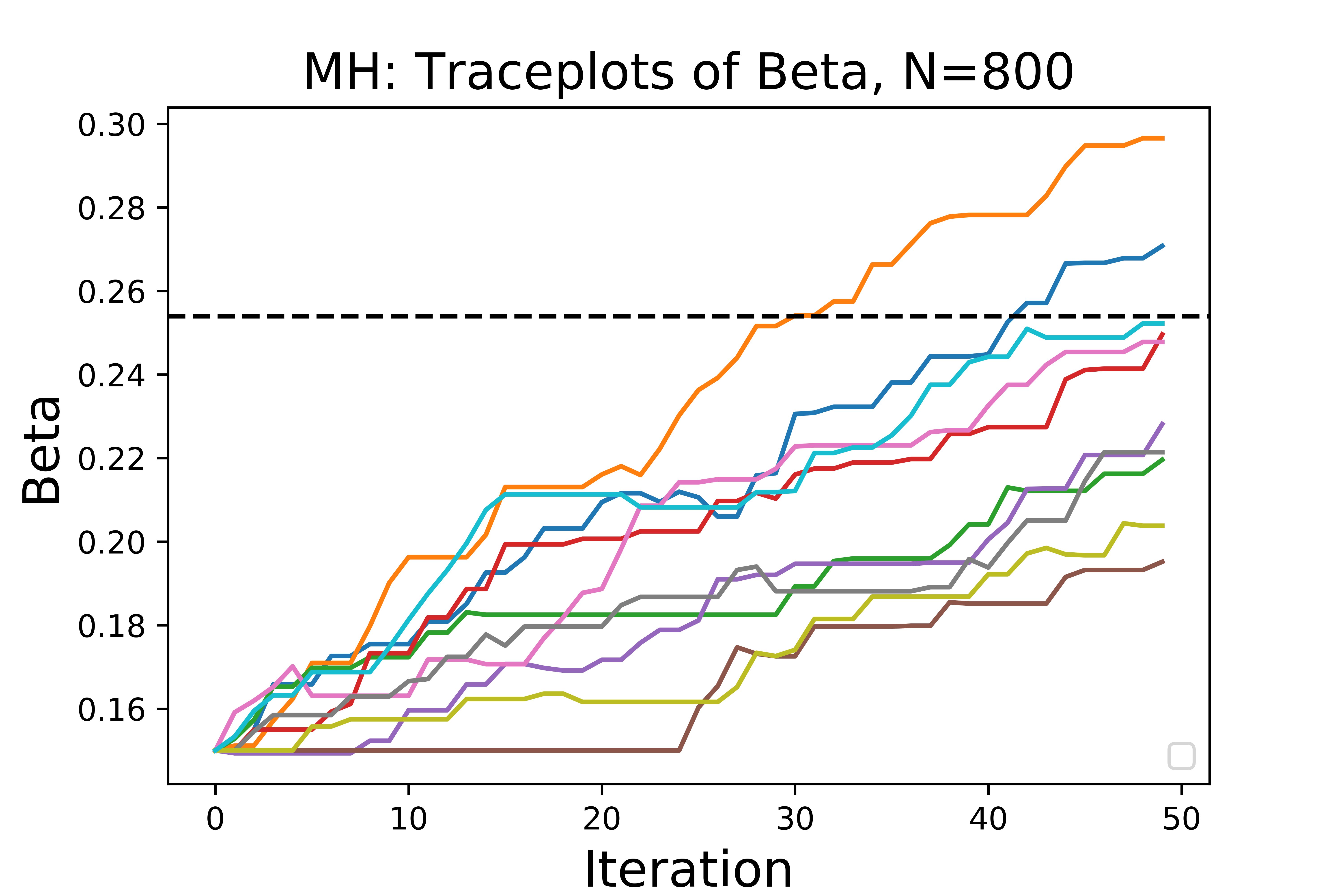}}
\hfill
\subfigure[]{\includegraphics[width=0.48\linewidth]{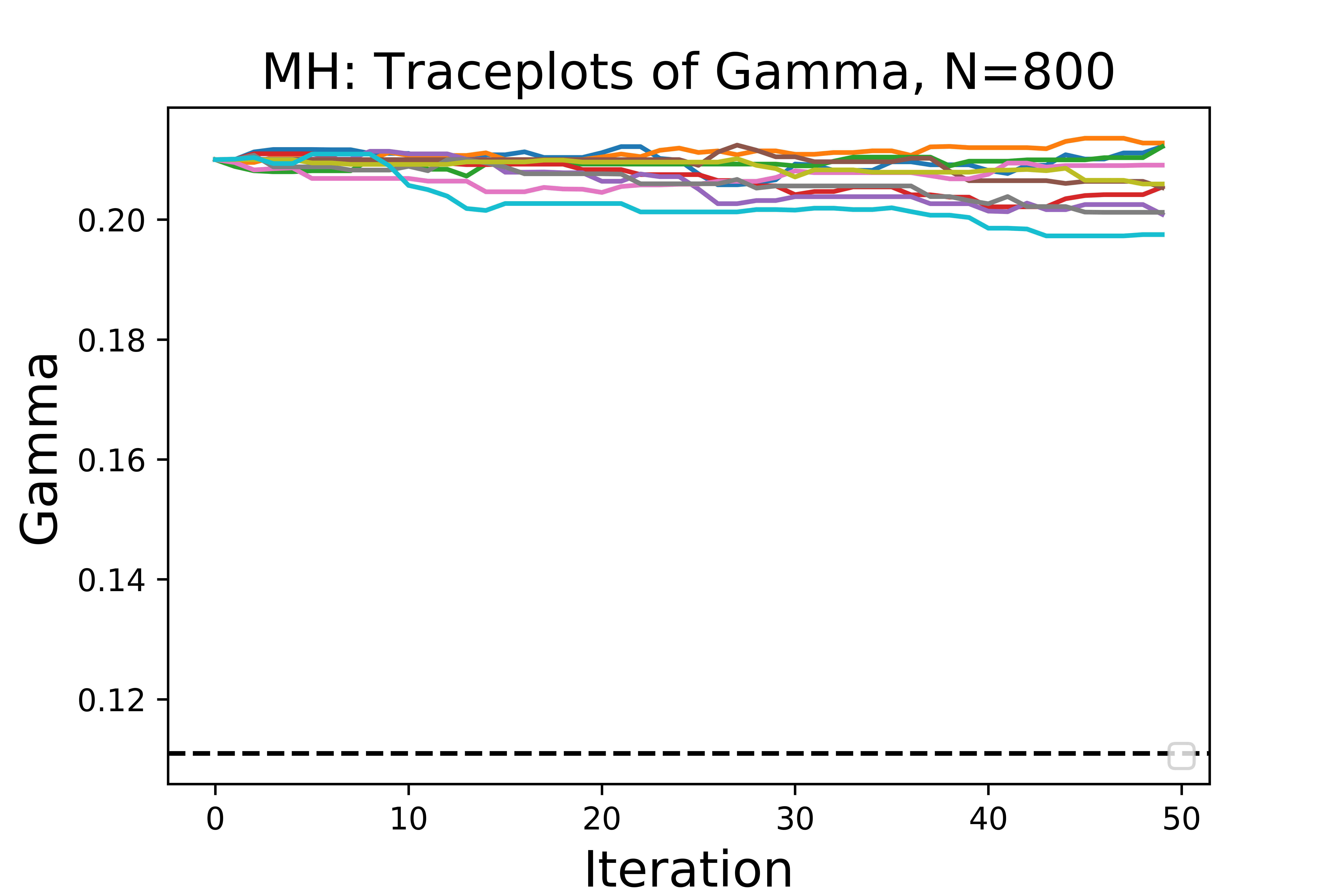}}
\hfill
\subfigure[]{\includegraphics[width=0.48\linewidth]{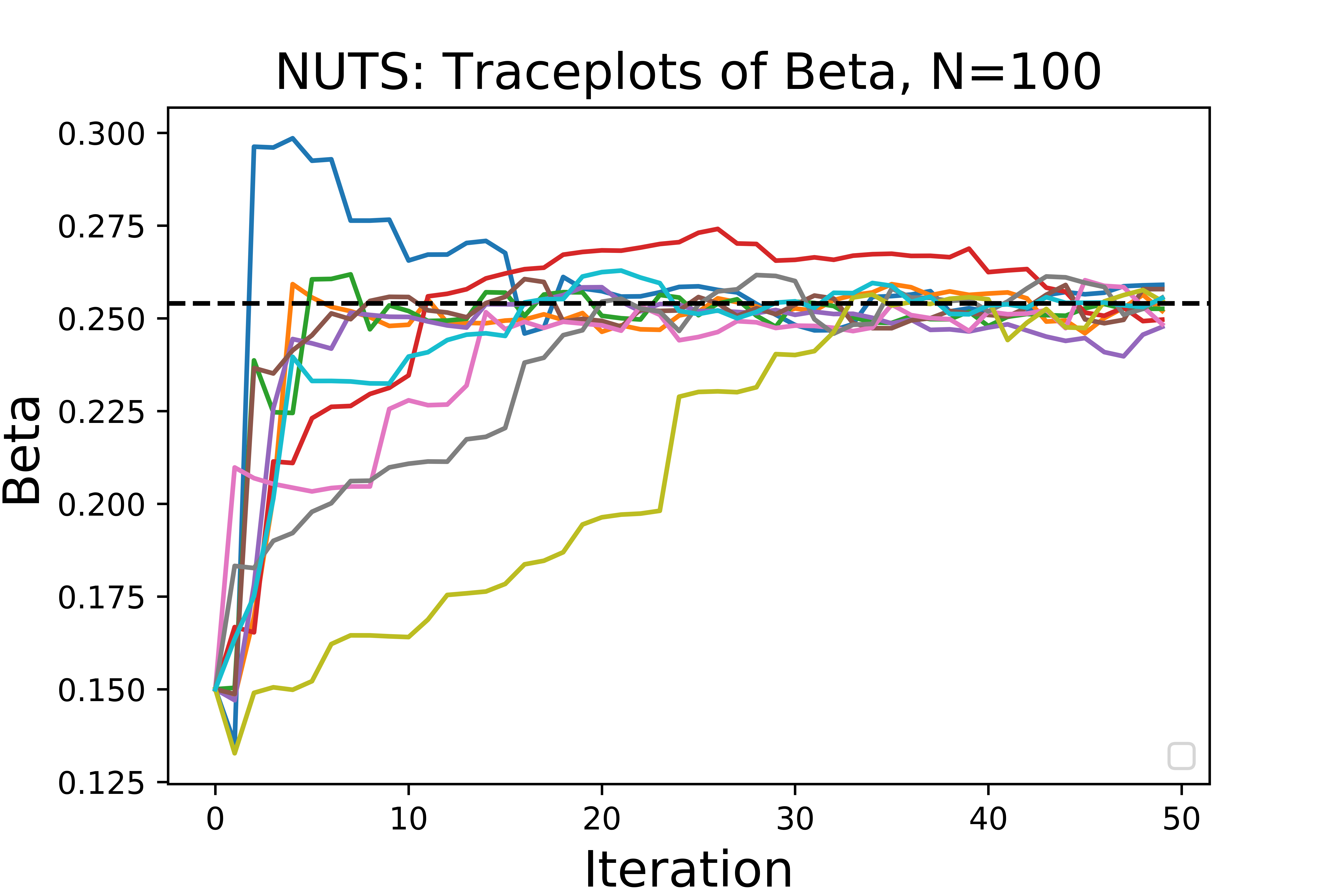}}
\hfill
\subfigure[]{\includegraphics[width=0.48\linewidth]{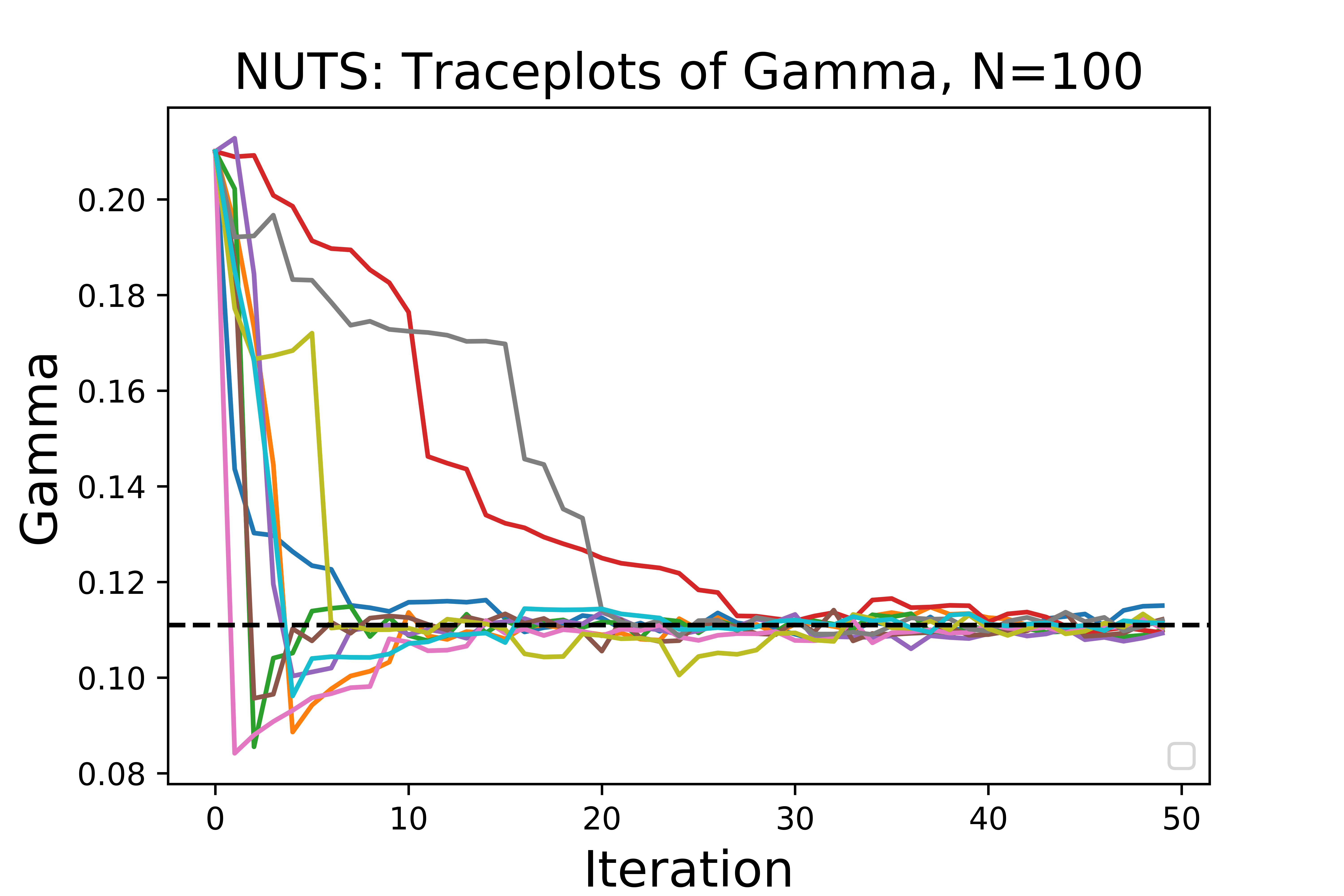}}
\caption{10 independent Markov chains with the same initial starting point but different random number seeds. Traceplots for Beta and Gamma using MH (a)-(b) and NUTS (c)-(d) on the top and bottom rows, respectively. The dashed line on all plots are the true values. We present results for MH and NUTS that have similar computation times (see Table \ref{table:MHvsNUTS}).}
\label{fig:covergence_rates}
\end{figure}

We run the particle filter to obtain estimates of the log-likelihood and the gradient of the log-likelihood w.r.t $\beta$ across a range of 100 values of $\beta$ equally spaced from 0.248 and 0.260. The aim is to maximise the log-likelihood and have a gradient of 0 when $\beta=0.254$. The results can be seen in Figure \ref{EpidemicCurves} (c) and (d). The Mean Squared Error (MSE) between the true and inferred values of $\theta$ for both MH and NUTS as well as the time taken in seconds when using different numbers of particles, $N$, can be seen in Table \ref{table:MHvsNUTS}. The MSE and computational time are averaged over 10 runs, which use different random number seeds for 50 MCMC iterations. Using 10 different random number seeds will produce different realisations of when resampling occurs (as explained in Section \ref{sec:CRNresampling}) so in turn will converge to the true values of $\theta$ at different rates. Averaging over the 10 runs gives a more accurate representation of the typical convergence accuracy and the time taken. This is exemplified in Figure \ref{fig:covergence_rates} when using M-H (subplots: a and b) and NUTS (subplots: c and d) for the SIR model. Each line is an independent traceplot of an MCMC chain with the same initial starting values of 0.15 and 0.21 for $\beta$ and $\gamma$, respectively. Note for the MH proposal we use the same step-sizes for $\beta$ and $\gamma$ as are stated in \cite{sheinson2014comparison} of 0.005 and 0.001, respectively. It is evident when looking at Table \ref{table:MHvsNUTS} that using NUTS can minimise the MSE in less computation time when compared with MH.

\begin{table}[]
\centering
\caption{Comparison of the mean square error and timings between a Metropolis-Hastings and a NUTS proposal in p-MCMC for the SIR model in section \ref{model:SIR} for different numbers of particls, $N$.}
\begin{tabular}{c c c c}
\hline 
 &  $\beta$ & $\gamma$ & Time (secs)  \\
\hline 
N=50 & &  & \\
&  & &   \\ 
M-H & 0.0055  &  0.0095 &  2.3020 \\ 
NUTS & \bf{0.0010}  &  \bf{0.0006} &  25.4688 \\ 
\hline 
N=100 & &  & \\
&  & &   \\ 
M-H & 0.0051  &  0.0095 &  3.9031 \\ 
NUTS & \bf{0.0010}  &  \bf{0.0006} &  31.9671 \\ 
\hline 
N=200 & &  & \\
&  & &   \\ 
M-H & 0.0054  &  \bf{0.0094}&  7.6374 \\ 
NUTS & \bf{0.0009}  &  0.0097 &  51.1340 \\ 
\hline 
N=400 & &  & \\
&  & &   \\ 
M-H & 0.0049  &  0.0097 &  17.4208 \\ 
NUTS & \bf{0.0012}  &  \bf{0.0008} &   101.5174 \\ 
\hline 
N=800 & &  & \\
&  & &   \\ 
M-H & 0.0046  &  0.0094 &  39.4373 \\ 
NUTS & -  &  - &   - \\ 
\hline 

\end{tabular}
\label{table:MHvsNUTS}
\end{table}

\subsubsection{SEIR}\label{results_SEIR}

\begin{figure}[htp]
\centering
\subfigure[]{\includegraphics[width=0.48\linewidth]{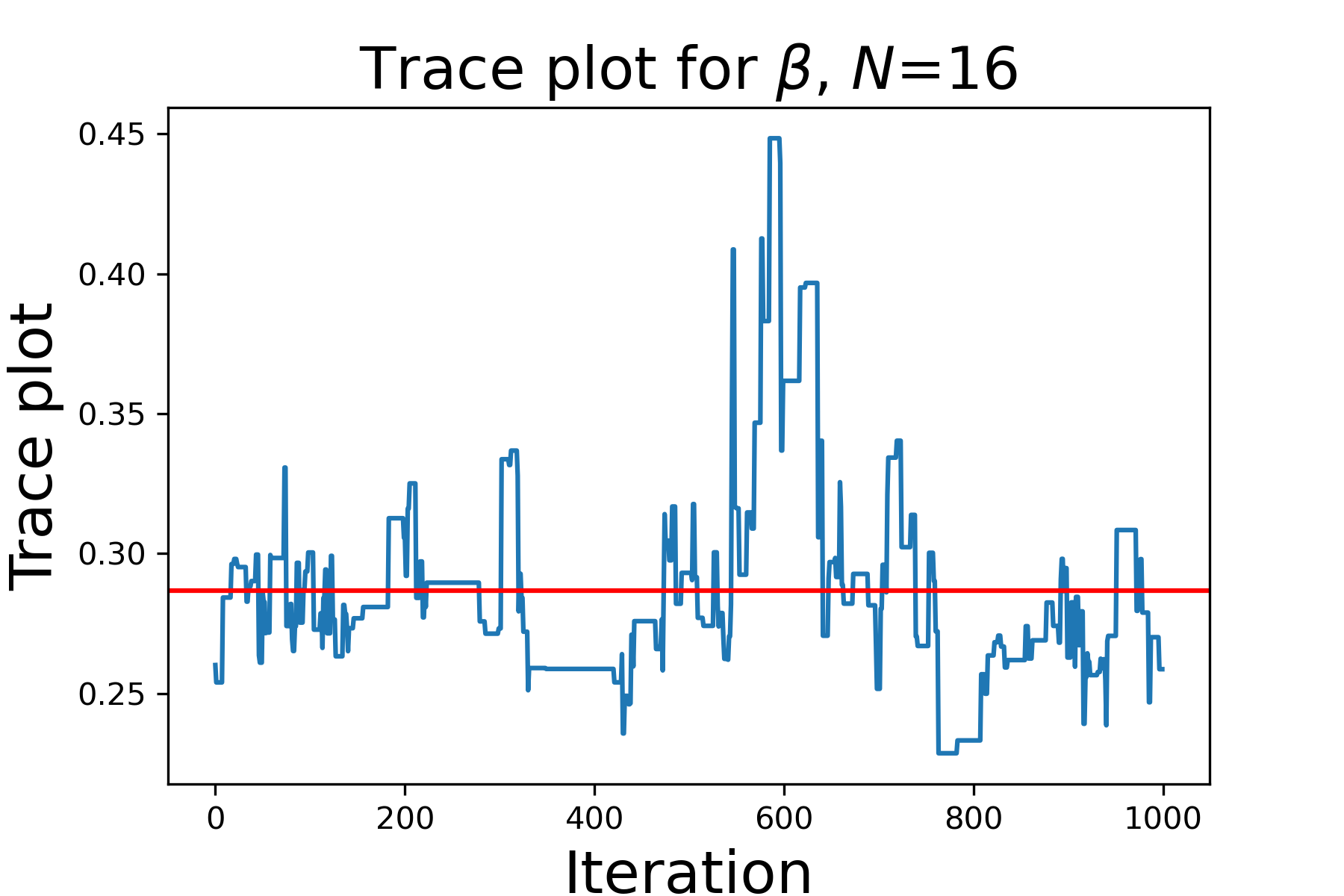}}
\hfill
\subfigure[]{\includegraphics[width=0.48\linewidth]{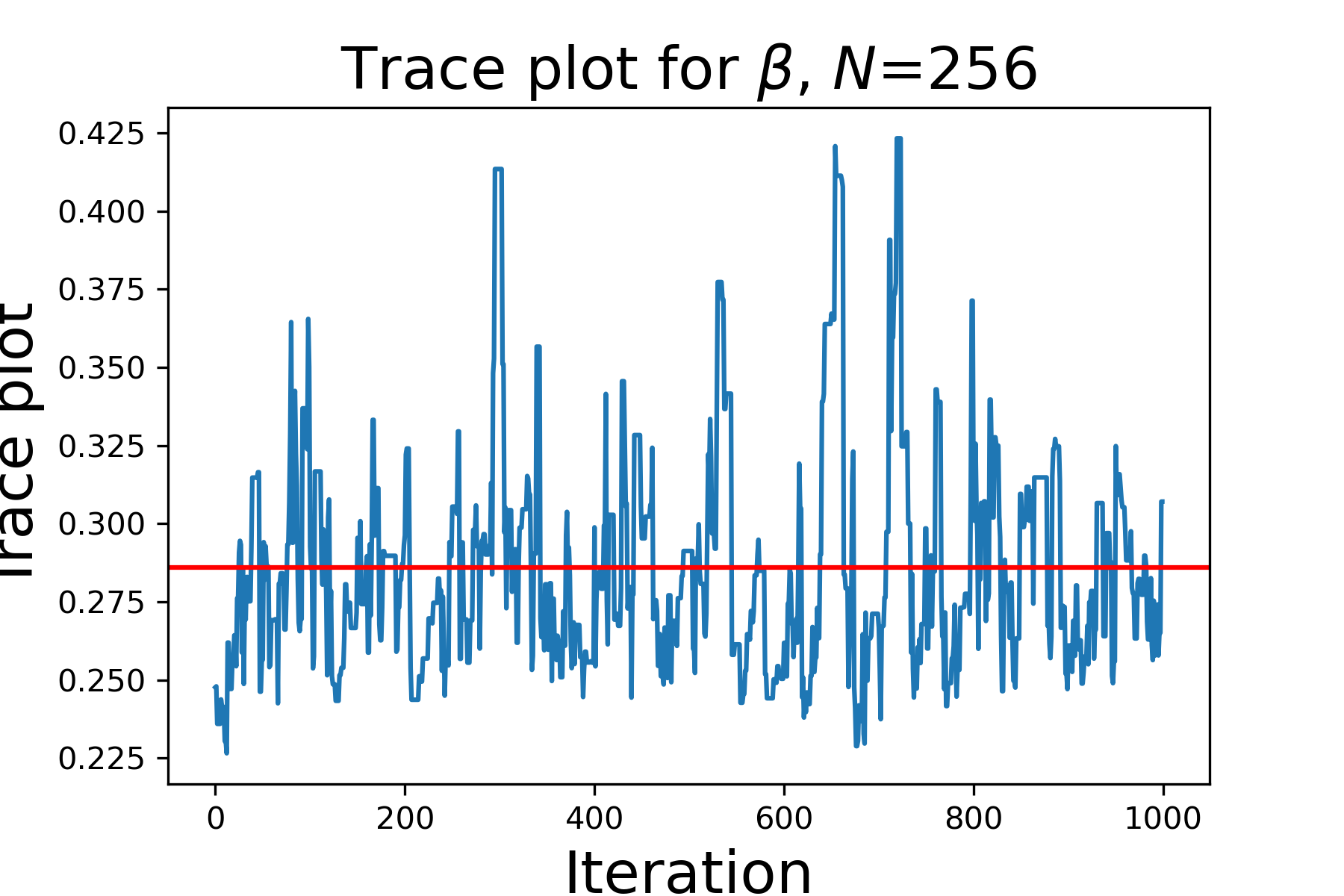}}
\hfill
\subfigure[]{\includegraphics[width=0.48\linewidth]{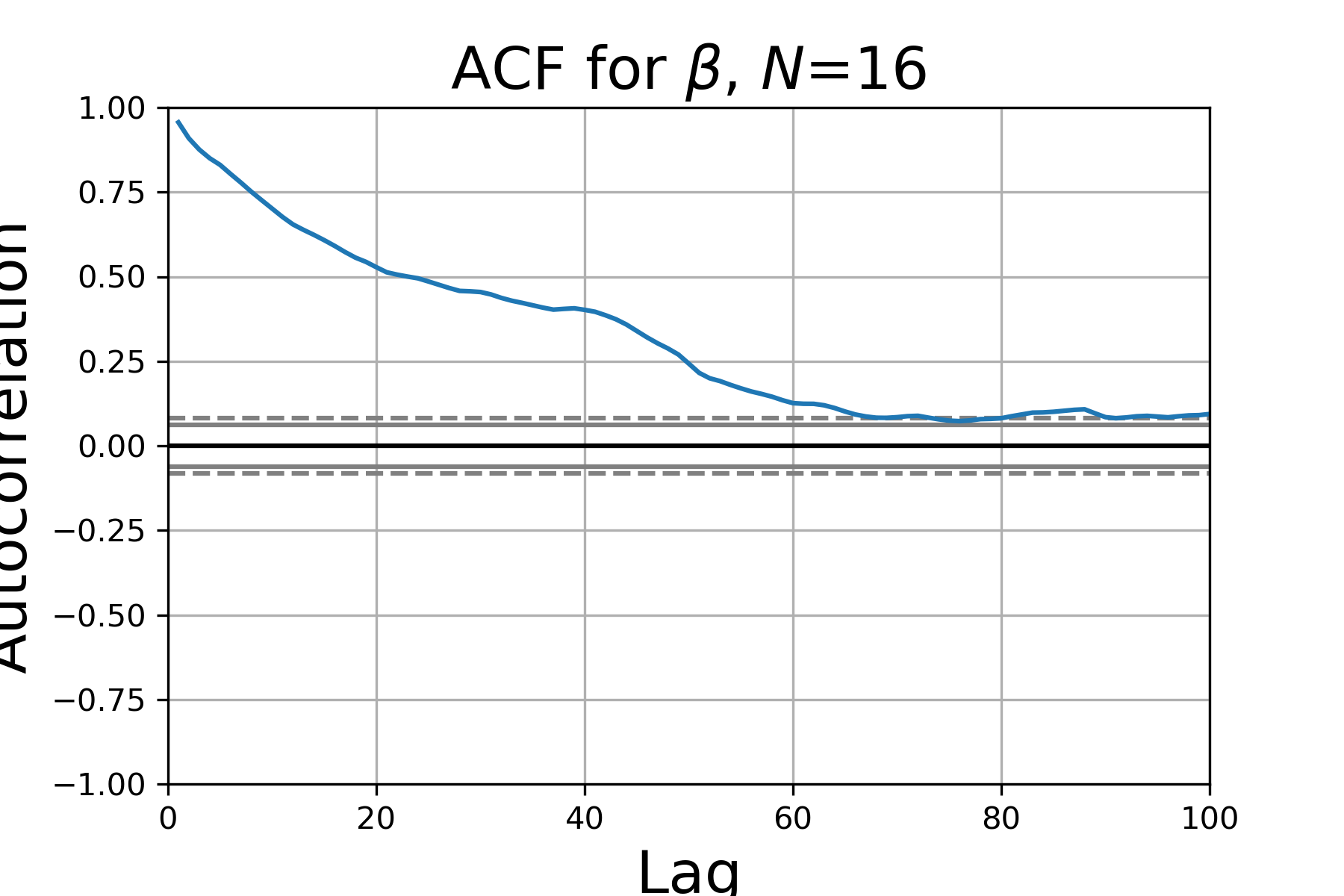}}
\hfill
\subfigure[]{\includegraphics[width=0.48\linewidth]{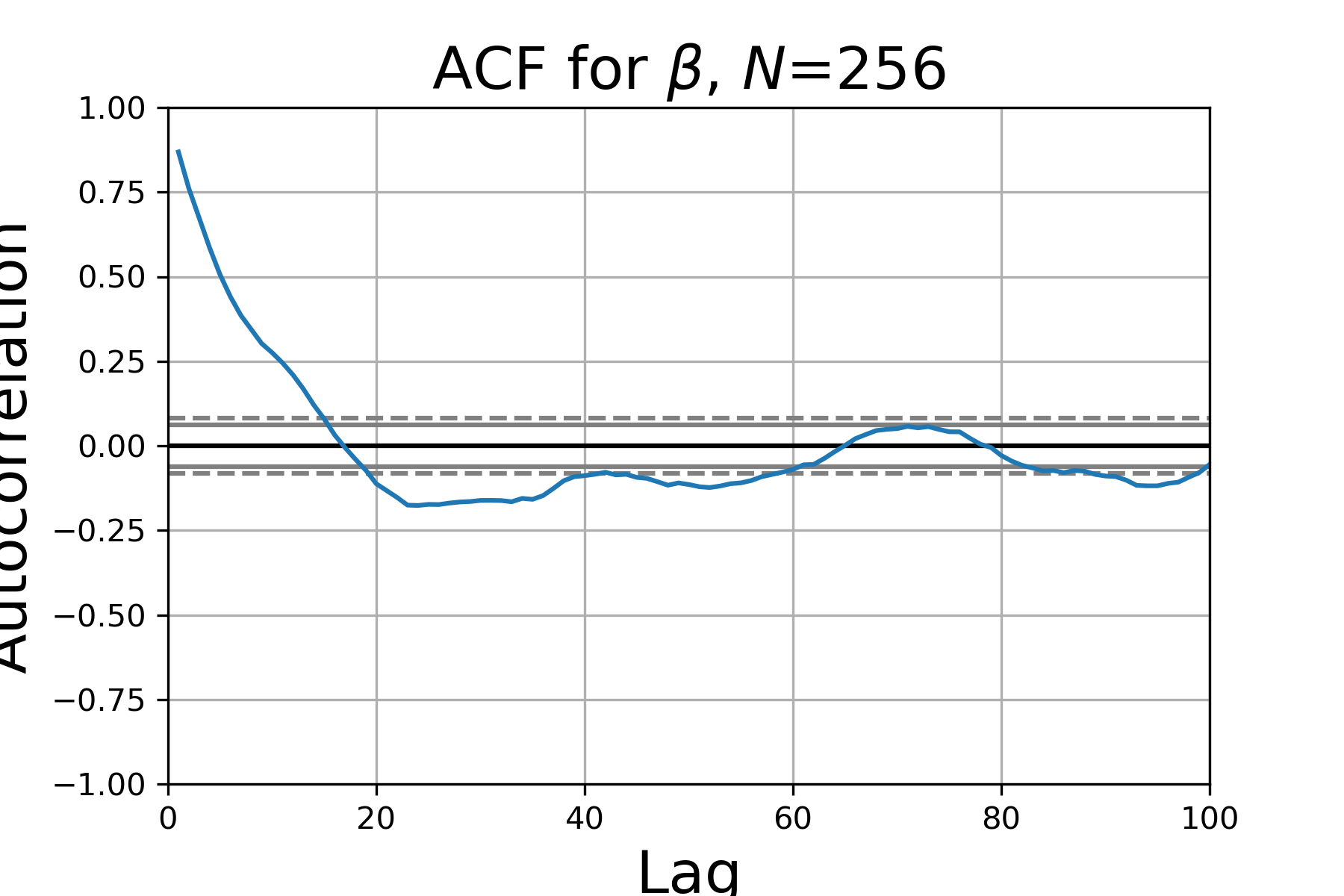}}
\caption{Convergence plots for $\beta$ in the SEIR model in \ref{model:SEIR} when using NUTS. The corresponding trace plots (a)-(b) and ACF plots (c)-(d) when $N=$ 16 and 256 in table \ref{table:convergence_SEIR}.}
\label{fig:SEIR_convergence}
\end{figure}

Secondly we show how well the Markov-chain converges to the true parameters when running with different numbers of particles, $N$. Table \ref{table:convergence_SEIR} outlines the results for the SEIR model in section \ref{model:SEIR} when using NUTS with $\Delta$ set to 0.0055. We run the simulation for 2000 samples with the first 1000 discarded as burn-in. The true values of $\beta$, $\gamma$ and $\delta$ set to 0.254, 0.111 and 0.4, respectively. Figure \ref{fig:SEIR_convergence} (a) and (b) shows the trace plots for $\beta$ when $N=$ 16 and $N=$ 256, respectively. By eye it is evident that using $N=$ 256 results in a Markov-chain that has better mixing which in turn explores $\pi(\theta)$ more efficiently than having $N=$ 16. This is backed up by looking at the acceptance rate in Table \ref{table:convergence_SEIR} with $N=$ 256 resulting in 0.53 of samples being accepted compared with 0.20 for $N=$ 16. The auto-correlation function (ACF) plots when having $N=$ 16 and $N=$ 256 can be seen in Figures \ref{fig:SEIR_convergence} (c) and (d), respectively. ACF plots shows the auto-correlation between samples in the Markov-chain as a function of a user-specified lag which we select to be 100. A plot that trends towards 0 in fewer lags is seen to be better, which is the case when $N=$ 256. Table \ref{table:convergence_SEIR} outlines the integrated auto-correlation time (IACT) for each parameter. This is a numerical measure of the area under the ACF plot and a lower value represents less correlation between consecutive samples.

The effective sample size (ESS) for different numbers of particles is also shown in Table \ref{table:convergence_SEIR} as well as the time taken in seconds the p-MCMC takes to run. This gives an indication of the number of independent samples it would take to have the same estimation power as a set of auto-correlated samples. The ESS/time taken in seconds shows how efficient the algorithm is by calculating how many independent samples are drawn per second. 

\begin{figure}[htp]
\centering
\subfigure[]{\includegraphics[width=0.48\linewidth]{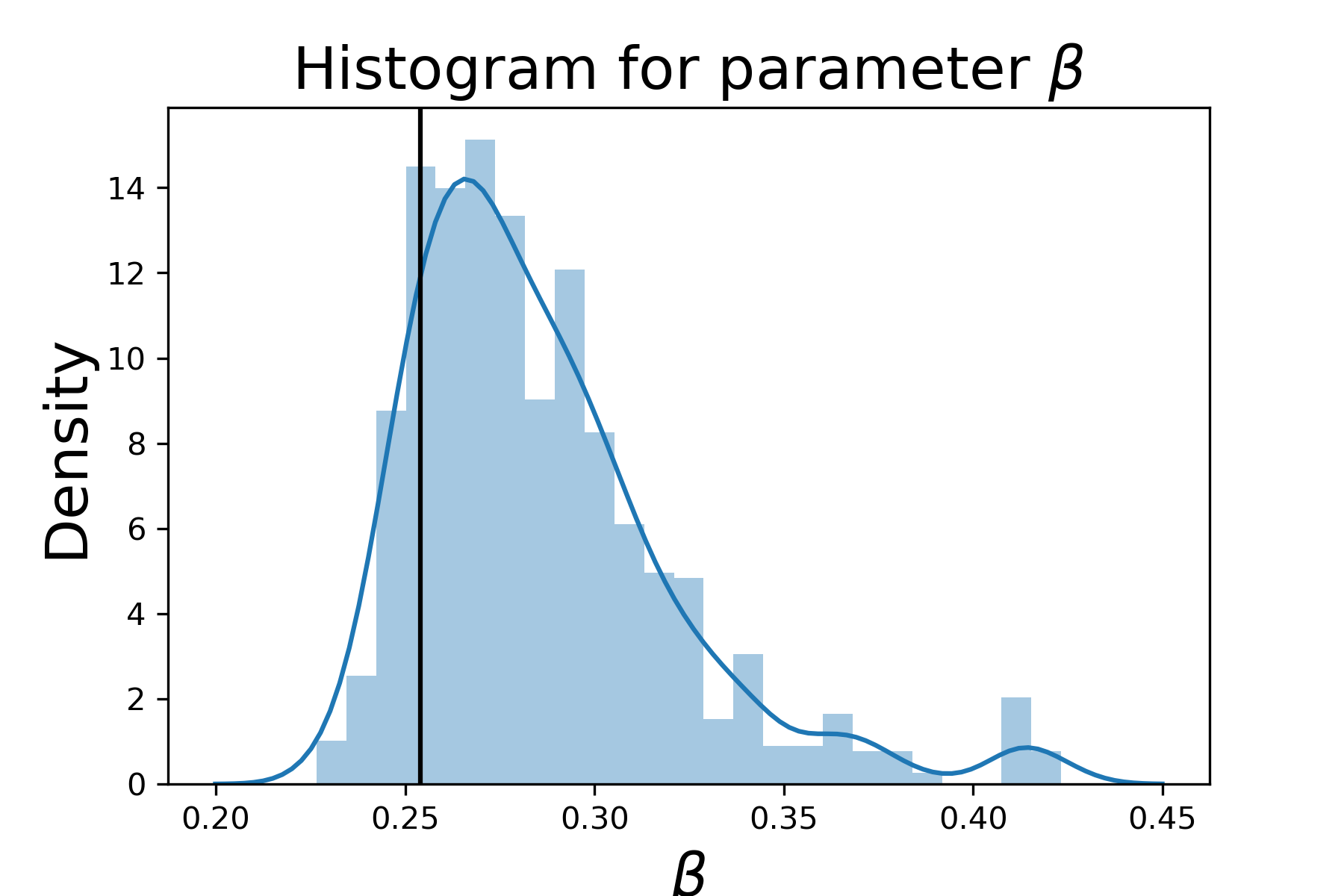}}
\hfill
\subfigure[]{\includegraphics[width=0.48\linewidth]{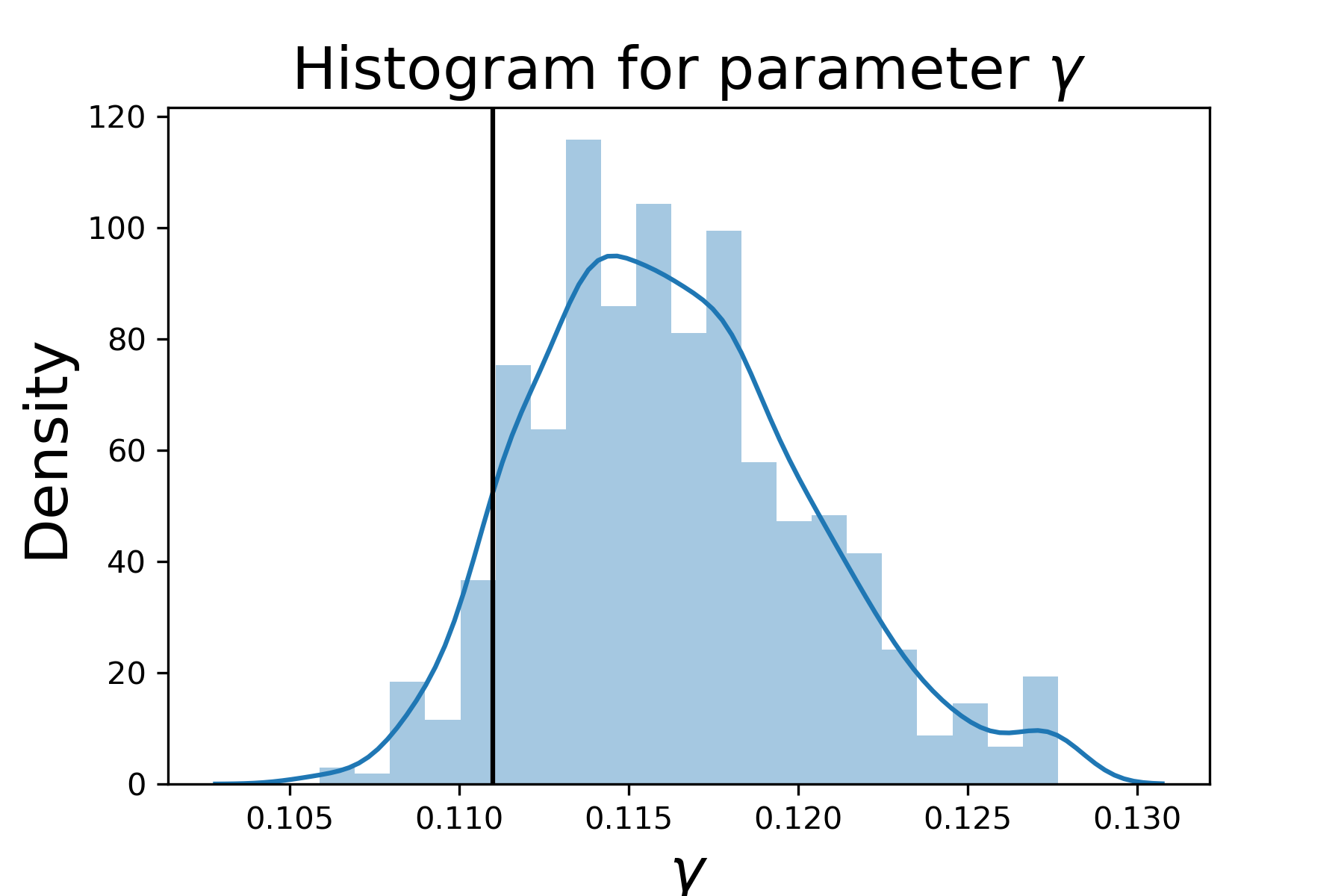}}
\hfill
\subfigure[]{\includegraphics[width=0.48\linewidth]{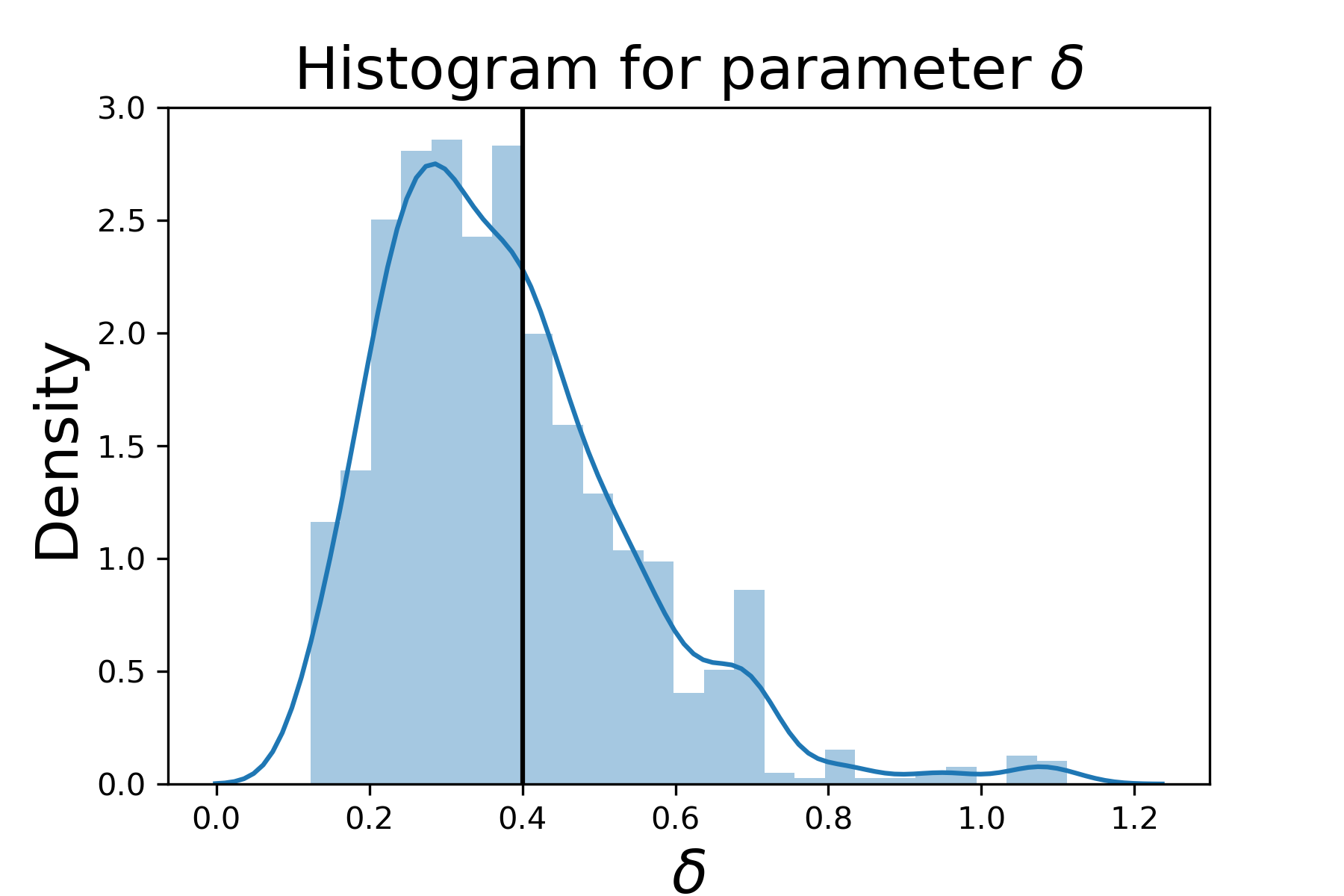}}
\hfill
\subfigure[]{\includegraphics[width=0.48\linewidth]{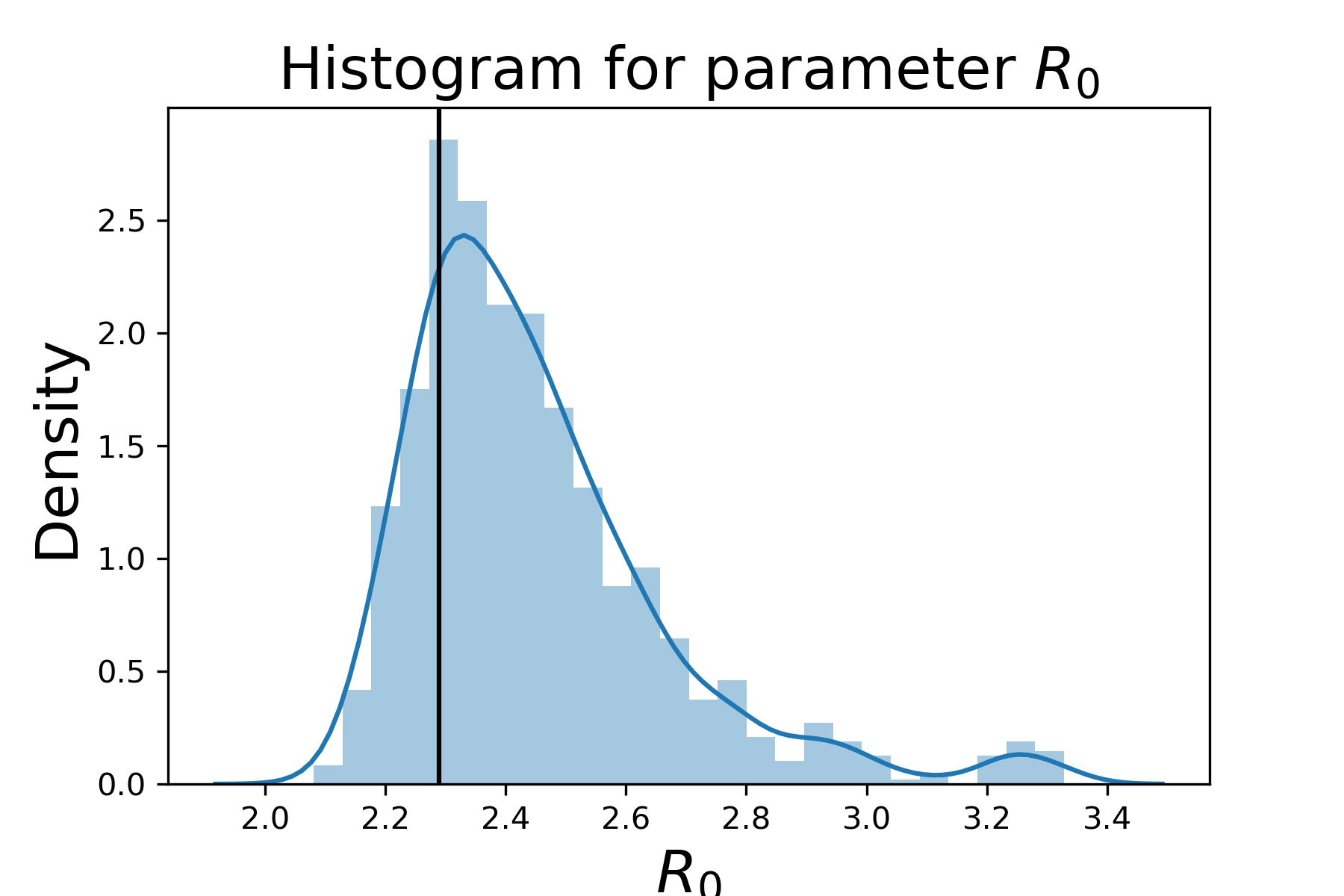}}
\caption{The marginal distribution plots for $\beta$, $\gamma$, $\delta$ and $R_t$ in (a) (b) (c) (d), respectively, when using NUTS and $N=$ 256 for the SEIR model in \ref{model:SEIR}.}
\label{fig:SEIR_histograms}
\end{figure}

The predicted marginal distributions, when using $N=$ 256 and NUTS, for the parameters $\beta$, $\gamma$ and $\delta$ can be seen in Figures \ref{fig:SEIR_histograms} (a), (b) and (c), respectively. We also show the predicted distribution around the the basic reproduction number, $R_0$, in Figure \ref{fig:SEIR_histograms} (d). This is calculated by $R_0 = \beta/\gamma$.

\begin{table}[]
\caption{Convergence statistics of a p-MCMC chain when inferring parameters of the SEIR model in section \ref{results_SEIR} when using NUTS and different values of particles, $N$.}
\begin{tabular}{clllll}
                                        \\\hline
\multicolumn{1}{c}{\textbf{Num. Particles}}     & 16 & 32 & 64 & 128 & 256 \\ \hline

\multicolumn{1}{c}{\textbf{Mean Estimate}} &    &    &    &     &         \\
\multicolumn{1}{c}{$\beta$}       &  0.287  & 0.273   &  0.293  &  0.299   &   0.286      \\
\multicolumn{1}{c}{$\gamma$}      &  0.116  &  0.115  &  0.117  &  0.118   &  0.116       \\
\multicolumn{1}{c}{$\delta$}      & 0.368   & 0.429   &  0.337  &  0.321   &  0.379       \\ \hline

\multicolumn{1}{c}{\textbf{IACT}}          &    &    &    &     &         \\
\multicolumn{1}{c}{$\beta$}       &  66  &  16  &  35  &  17   &  2       \\
\multicolumn{1}{c}{$\gamma$}      &  65  &  11  &  35  &  15   &  2       \\
\multicolumn{1}{c}{$\delta$}      &  69  &  17  &  26  &  9   &  7   
\\\hline
\multicolumn{1}{c}{\textbf{ESS}}          &    &    &    &     &         \\
\multicolumn{1}{c}{$\beta$}       & 16   & 46   & 21   &   65 &  81       \\
\multicolumn{1}{c}{$\gamma$}      &  15  & 50   & 11   &  74   &  87       \\
\multicolumn{1}{c}{$\delta$}      &  13  &  38  &  30  &  63   &  73       \\ \hline
\multicolumn{1}{c}{\textbf{Time (s)}}      &  5952  &  7798  & 9985   &  12831   &  21763       \\ \hline
\multicolumn{1}{c}{\textbf{ESS/s}}          &    &    &    &     &         \\
\multicolumn{1}{c}{$\beta$}       & 0.003   & 0.006   & 0.002   &  0.005 &  0.004       \\
\multicolumn{1}{c}{$\gamma$}      &  0.003  & 0.006   & 0.001   &  0.006   &  0.004       \\
\multicolumn{1}{c}{$\delta$}      &  0.002  &  0.005  &  0.003  &  0.005   &  0.003       \\ \hline
\multicolumn{1}{c}{\textbf{Acc. Rate}}     &  0.20  &  0.33  &  0.46  & 0.52    & 0.53       \\ \hline
\end{tabular}
\label{table:convergence_SEIR}
\end{table}

\section{Conclusions and Future Work}\label{sec:conclusions}

We have shown that using NUTS as the proposal can improve accuracy when inferring the parameters of a simple SIR model over the standard MH proposal. We have also outlined multiple methods to assess convergence of MCMC chains. These are a necessity to ascertain if the results obtained from the p-MCMC algorithm are useful.

Another popular method for modelling epidemics described by Reed-Frost assumes that people becoming infected with a disease follows a Binomial distribution. We do not include this model in our analysis due to (\ref{eqn:dxk}) being non-differentiable when the particle state equations follow a discrete distribution. A fruitful direction for future work would involve including the ideas described in \cite{diffPF_OptimalTransport}, which leverages optimal transport ideas to mimic the multinomial distribution, and mimic the Binomial distribution. 

In this piece of work we have only considered simulated data to provide a proof of concept. A next step would be to apply these methods to more sophisticated epidemiological models, as seen in \cite{moore2021refining}, and to use real data pertinent to disease outbreaks. 

One last direction for future work could involve ingesting multiple data streams with different reporting latency. It has been shown in \cite{rosato2021fusing} that certain combinations of datasets can improve accuracy when predicting COVID-19 related deaths.

\section*{Acknowledgment}

The authors would like to acknowledge Lee Devlin and the UK Health Security Agency (UKHSA) for their support.

\bibliographystyle{ieeetr}
\bibliography{bibliography.bib}
\vspace{12pt}
\appendices

\section{Derivatives of multivariate log normal}	\label{app:normalderiv}

If
\begin{align}
	\mathcal{N}(x; \mu, C) & = \frac{\exp\left(-\frac12(x - \mu)^TC^{-1}(x - \mu)\right)}{\sqrt{|2\pi C|}}
\end{align}
then
\begin{align}
	\frac{\partial}{\partial x}\log\mathcal{N}  & = -C^{-1}(x - \mu) \label{eqn:dlogNdx} \\
	\frac{\partial}{\partial \mu}\log\mathcal{N} & = C^{-1}(x - \mu) \label{eqn:dlogNdmu} \\
	\frac{\partial}{\partial C}\log\mathcal{N} & = -\frac12\left(C^{-1} - C^{-1}(x - \mu)(x - \mu)^TC^{-1}\right). \label{eqn:dlogNC}
\end{align}

\begin{algorithm}[h]
	\caption{\label{p-mcmcMH} Pseudocode for p-MCMC using a MH Proposal for model \ref{model:SIR} when inferring $\beta$ and $\gamma$}
	\begin{algorithmic}[1]
		\REQUIRE $\theta_0$   \COMMENT{Inputs}
		\STATE Run Algorithm~\ref{app:particleFilter} with $\theta=[\beta, \gamma]$ to obtain estimate of $\log p(y_{1:T} | \theta)$
		\FOR{i = 1, \dots, M \do}
		    \STATE Sample $\theta'=[\beta', \gamma']$ from a proposal distribution $q(\theta'|\theta)$
		    \STATE Run Algorithm~\ref{app:particleFilter} with $\theta'$ to obtain estimates of $\log p(y_{1:T} | \theta')$
		    \STATE Compute acceptance probability, $\alpha(\theta,\theta')$ using \eqref{eq:MHalgorithm_}
		    \STATE Sample $u\sim U[0,1]$
		    \IF{$u < \alpha(\theta,\theta')$}
		        \STATE accept $\theta'$, $\log p(y_{1:T} | \theta')$
		    \ELSE
		        \STATE accept $\theta$,  $\log p(y_{1:T} | \theta)$
		    \ENDIF
		\ENDFOR
	\end{algorithmic}
\end{algorithm}

\begin{algorithm}[h]
	\caption {\label{app:particleFilter}Pseudocode for the Particle Filter}
	\begin{algorithmic}[1]
		\REQUIRE $\theta$, $Y_{1:T}$, $N$   \COMMENT{Inputs}
		\STATE Initialise $x_{0}^i$, $dx_{0}^i/d\theta$, $\log(w_{0}^i)$, $d\log(w_{0}^i)/d\theta$
		\FOR{i = 1, \dots, T \do}
		    \STATE If $N_{eff}$ (\ref{eq:neff}) $\leq$ threshold:
		    
		        Resample $x_{t-1}^i$, $\log(w_{1:k-1}^i)$, $dx_{t-1}^i/d\theta$, $d\log(w_{1:t-1}^i)/d\theta$\;
		    \STATE Sample the new particles $x_t^i$ and calculate the particle derivatives using (\ref{eqn:dxk}).\;
		    \STATE Calculate the particle weights and derivatives of particle weights using (\ref{weightupdate}) and (\ref{simplifieddweightupdatedtheta}), respectively.
		\ENDFOR
	\end{algorithmic}
\end{algorithm}

\begin{algorithm}[h]
	\caption{\label{particleFilter} Pseudocode for p-MCMC using NUTS as the proposal.}
	\begin{algorithmic}[1] 
	    \REQUIRE $\ell(\theta)$, $\nabla\mathcal{L}(\theta)$ = Run Algorithm 2
		\STATE Initialise $x_{0}^i$, $dx_{0}^i/d\theta$, $\log(w_{0}^i)$, $d\log(w_{0}^i)/d\theta$
			\FOR{i = 1, \dots, M \do}
		    \STATE NUTS = Algorithm 3 in \cite{hoffman2014no}
		\ENDFOR
	\end{algorithmic}
\end{algorithm}

\end{document}